\documentclass{JHEP3}
\usepackage{amsmath}
\usepackage{axodraw}
\usepackage{epsfig}
\usepackage{amssymb}
\usepackage{mathrsfs}
\usepackage{mathcomp}
\usepackage{cite}
\usepackage{stmaryrd}
\usepackage[latin1]{inputenc}

\usepackage{cite}
\usepackage{epsfig}
\usepackage{xspace}
\usepackage{graphics}
\usepackage[latin1]{inputenc}

\def\hide#1{}

\newcommand{\madevent}{M\scalebox{0.8}{AD}E\scalebox{0.8}{VENT}\xspace}

\newcommand{\herwig}{\scalebox{0.8}{HERWIG}\xspace}

\newcommand{\sherpa}{S\scalebox{0.8}{HERPA}\xspace}
\newcommand{\diclus}{D\scalebox{0.8}{ICLUS}\xspace}
\newcommand{\pythia}{P\scalebox{0.8}{YTHIA}\xspace}
\newcommand{\jetset}{J\scalebox{0.8}{ETSET}\xspace}

\newcommand{\ariadne}{A\scalebox{0.8}{RIADNE}\xspace}

\newcommand{\as}{\ensuremath{\alpha_{\mathrm{s}}}}

\newcommand{\kT}{\ensuremath{k_{\perp}}}
\newcommand{\pT}{\ensuremath{p_{\perp}}}

\newcommand{\ECM}{\ensuremath{E_{\mathrm{CM}}}}

\newcommand{\particle}[1]{\ensuremath{\mathrm{#1}}}

\newcommand{\el}{\particle{e}}

\newcommand{\ee}{\ensuremath{\el^+\el^-}}

\def\mrm#1{\mathrm{#1}}

\def\sub#1{\ensuremath{_{\mrm{#1}}}}
\def\sup#1{\ensuremath{^{\mrm{#1}}}}

\def\f2d3{\ensuremath{F_2^{\mrm{D}3}}}

\def\done#1{}

\providecommand{\eqref}[1]{eq.~(\ref{#1})\xspace}
\renewcommand{\eqref}[1]{eq.~(\ref{#1})\xspace}

\newcounter{aenumct}

\newcounter{ienumct}

\renewenvironment{itemize}{\begin{list}{$\bullet$}%
{\setlength{\topsep}{0mm}\setlength{\partopsep}{1mm}%
\setlength{\itemsep}{0mm}\setlength{\parsep}{1mm}}}%
{\end{list}}

\newcounter{enumct}
\renewenvironment{enumerate}{\begin{list}{\arabic{enumct}.}%
{\usecounter{enumct}\setlength{\topsep}{1mm}%
\setlength{\partopsep}{1mm}\setlength{\itemsep}{0mm}%
\setlength{\parsep}{1mm}}}{\end{list}}

\def\eg{\emph{e.g.}}

\def\asme{\alpha_{\mrm{sME}}}
\def\yms{\ensuremath{y_{\mrm{MS}}}}

\def\Qcut{\ensuremath{Q_{\mrm{cut}}}}
\def\qms{\ensuremath{q_{\mrm{MS}}}}
\def\Qms{\ensuremath{Q_{\mrm{MS}}}}
\def\dms{\ensuremath{d_{\mrm{MS}}}}
\def\rms{\ensuremath{\rho_{\mrm{MS}}}}
\def\rmax{\ensuremath{\rho_{\mrm{max}}}}

\def\dmax{\ensuremath{d_{\mrm{max}}}}
\def\sme{\sigma^{\mrm{ME}}}
\def\sps{\sigma^{\mrm{PS}}}
\def\smeps{\sigma^{\mrm{ME+PS}}}
\def\gme{\Gamma^{\mrm{ME}}}
\def\gps{\Gamma^{\mrm{PS}}}
\def\evol#1#2{f^{{#1}\shortrightarrow{#2}}}

\newcommand{\alpgen}{A\scalebox{0.8}{LPGEN}\xspace}
\newcommand{\helac}{H\scalebox{0.8}{ELAC}\xspace}
\newcommand{\luclus}{L\scalebox{0.8}{UCLUS}\xspace}

\preprint{LU-TP 07-38}

\title{Merging parton showers and matrix elements\\ --- back to
  basics\footnote{Work supported in part by the Marie Curie RTN
    ``MCnet'' (contract number MRTN-CT-2006-035606).}}

\author{Nils Lavesson and Leif Lönnblad\\
  Dept.~of Theoretical Physics,
  Sölvegatan 14A, S-223 62  Lund, Sweden\\
  E-mail: \email{Nils.Lavesson@thep.lu.se}
    and \email{Leif.Lonnblad@thep.lu.se}}

  \abstract{
    We make a thorough comparison between different schemes of merging
    fixed-order tree-level matrix element generators with
    parton-shower models. We use the most basic benchmark of the
    $\mathcal{O}(\alpha_S)$ correction to $e^+e^-\to\mbox{jets}$, where
    the simple kinematics allows us to study in detail the transition
    between the matrix-element and parton-shower regions. We find that
    the CKKW-based schemes give a reasonably smooth transition between
    these regions, although problems may occur if the parton shower
    used is not ordered in transverse momentum. However, the so-called
    Pseudo-Shower and MLM schemes turn out to have potentially serious
    problems due to different scale definitions in different regions of
    phase space, and due to sensitivity to the details in the initial
    conditions of the parton shower programs used.

}

\begin{document}

\sloppy

\section{Introduction}
\label{sec:intro}

Accurate simulations of multi-jet final states are important for
current experiments and will become even more so once the LHC
starts. At the LHC the production rate for such states will be large
due to the huge available phase space. Hadronic multi-jet states are
used for many of the discovery channels for new physics and the main
irreducible background comes from QCD. A good theoretical
understanding and accurate physics simulations of multi-jet QCD states
are therefore essential tools for understanding and analyzing LHC
data.

To be able to compare the predictions of a model with a collider
experiment, a description of final state hadrons is needed. There are a
few phenomenological models available to describe the production of
hadrons, but they all require that the perturbative emissions are well
described, especially in the soft and collinear regions, to give
reliable results. These collinear and soft emissions dominate the
multi-parton cross section and can be taken into account to all
orders, if one approximates the emissions to be strongly ordered, as
is done in parton shower models. However, when the strong ordering no
longer holds, which is the case if we have several hard and widely
separated jets, the parton shower models become unreliable.

In order to improve the description of multi-jet states, full matrix
elements can be used. These describe the process correctly up to a
given order in the strong coupling constant. However, the matrix
elements become difficult to calculate for high parton multiplicities
or if one goes beyond tree-level. They also contain divergences in
the soft and collinear limits and need to be regulated using cutoffs.

The idea behind merging algorithms is to let the matrix elements
describe the hard emissions and use the parton shower to describe the
soft and collinear emissions. To accommodate this, the phase space needs
to be split into two well defined regions, one where emissions are
generated by tree-level matrix elements, and one where emissions are
generated by the parton shower. To avoid double counting and dead
regions, the two regions should have no overlaps and together cover the
entire phase space. The scale that describes the border between the
regions is called the merging scale.

Some extra care is needed to avoid an artificial dependence on the
merging scale. The matrix elements contain divergences in the soft and
collinear regions, whereas the emission probability in these regions in
the parton shower is finite since it is regulated by Sudakov form
factors. The effects of the Sudakov form factors need to be included in
the matrix element part of the phase space as well, making the state
from the matrix element exclusive. Also a running coupling similar to
that of the shower needs to be introduced. If this is done correctly the
dependence on the merging scale should be minimal, however, especially
for corrections with several extra jets, a small residual dependence on
the merging scale from sub-leading terms is basically unavoidable.

There are four main algorithms that address the problem of merging
tree-level matrix elements and partons showers: CKKW
\cite{Catani:2001cc}, CKKW-L \cite{Lonnblad:2001iq}, Pseudo-Shower
\cite{Mrenna:2003if}, and MLM \cite{MLM}. CKKW was first published for
$e^+e^-$ collisions \cite{Catani:2001cc} and later extended to hadron
collisions \cite{Krauss:2002up}. Implementations have been made in
\sherpa \cite{Gleisberg:2003xi} and in \herwig \cite{Mrenna:2003if,
Corcella:2000bw}. CKKW has been used in several studies of vector boson
production in hadronic collisions \cite{Mrenna:2003if, Krauss:2004bs,
Krauss:2005nu, Gleisberg:2005qq, Alwall:2007fs}.  \makebox{CKKW-L} was also first
published for $e^+e^-$ \cite{Lonnblad:2001iq} and later extended to
hadrons and applied to $W$-production \cite{Lavesson:2005xu,
Alwall:2007fs}. All the CKKW-L results so far have been calculated using
the \ariadne \cite{ARIADNE92} implementation. The Pseudo-Shower scheme
was published in \cite{Mrenna:2003if}, where the algorithm was
described, and an implementation based on \pythia
\cite{Sjostrand:2003wg} was applied to both $e^+e^-$ and hadron
collisions. The MLM algorithm has been implemented in
\alpgen\cite{Mangano:2002ea}, \madevent\cite{Alwall:2007st} and
\helac\cite{Cafarella:2007pc}. There is also one implementation based on
\herwig and \madevent, which was used in \cite{Mrenna:2003if}. Several
studies using MLM have been performed for heavy quark and vector boson
production with incoming hadrons\cite{Mrenna:2003if, Mangano:2001xp,
Mangano:2006rw, Alwall:2007fs}. To our knowledge no calculation applying
MLM to $e^+e^-$ annihilation has been published.

Of all the implementations, only the \sherpa implementation of the CKKW
scheme and the \alpgen and \madevent implementations of MLM are publicly
available, while the others are obtainable from their respective
authours upon request.

There have been some assessments of the systematics of the various
algorithms done already, but the main focus so far has been the case
of hadron collisions. Although collisions with incoming hadrons
clearly are the most interesting in light of the upcoming LHC
experiments, they also include a lot of complications, such as
uncertainties from PDFs and BFKL-like corrections, that obfuscate the
basic properties of the merging algorithms.  $e^+e^-$ annihilation is
much simpler from a theoretical point of view and is therefore more
suitable for testing the basic properties of the different
algorithms. We believe it is essential to test the algorithms for
$e^+e^-$ annihilation before moving on to hadron collisions.

Systematics for $e^+e^-$ have been published for CKKW-L
\cite{Lonnblad:2001iq} and for CKKW and Pseudo-Shower
\cite{Mrenna:2003if}, but the systematics for the Pseudo-Shower approach
was quite limited. These studies are extended in this paper, where the
four algorithms listed earlier are applied to different parton shower
implementations. This allows us to thoroughly test which algorithms
live up to their promises.

To study systematics with as little complications as possible we only
look at the simple case of $e^+e^- \rightarrow q\bar{q}g$. In this
case the matrix element correction is already implemented in most of
the parton showers, using a simple reweighting of the hardest
splitting\footnote{We refer to these matrix element corrections
  of only the hardest splitting as \textit{reweighting} while the more
  general schemes on trial here are referred to as
  \textit{merging}.
}\cite{Gustafson:1988rq,Bengtsson:1987hr,Seymour:1995we,Seymour:1995df},
which makes this process particularly suitable for testing the
various algorithms. Ideally all the calculations should show small
deviations and yield more or less trivial results. However, we find
that this is not the case.

Since the process studied is rather simple, algorithms that perform well
should also be tested with more complicated processes before they can
be reliably used to predict experimental observables. Some of the
complications that can occur during the merging do not enter if only
first order corrections are applied. However, if an algorithm does not
perform well for this simple process, it is improbable that it will
work reliably for more complicated processes.

We note that besides the four merging algorithms presented here, there
are also other ways of combining matrix elements and parton showers,
\eg\ the methods based on modifyig the tree-level or NLO matrix
elements to \textit{match} the parton shower (see \eg\
\cite{Frixione:2002ik, Frixione:2006gn, Nason:2004rx, Nason:2006hfa,
Frixione:2007vw, Kramer:2005hw, Nagy:2005aa, Nagy:2007ty,
Giele:2007di}). Such
matching algorithms may also have a dependence on an artificial
matching scale. Although it may be interesting to also benchmark
these algorithms using the simplest $e^+e^-\to qg\bar{q}$ process, we
have not done so in this paper\footnote{After we wrote this paper a
  new algorithm was published \cite{Bauer:2008qh,Bauer:2008qj} which
  used a similar benchmarking.}.

In this paper we will concentrate on the behavior of the merging
schemes in absolute numbers. It would also be interesting to study
their formal properties in terms of leading double- and
single-logarithmic contributions to various cross sections, which so
far has only been done for the CKKW scheme\cite{Catani:2001cc}. We
plan to return to such issues in a future publication.

In this article we start by reviewing the theoretical aspects of the
four algorithms in section \ref{sec:theory}. Then we move on to showing
results from our implementations of the algorithms in section
\ref{sec:results} and finally in section \ref{sec:conclusions} we
present our conclusions.

\section{Theory}
\label{sec:theory}

All of the algorithms considered in this paper aim to do a good job of
merging matrix elements and partons showers. The issues that are
addressed are the same, namely to split the phase space in a clean
well defined way and to make the matrix element event exclusive by
introducing Sudakov form factors or using some other similar
suppression. The main aim is to minimize any artificial dependence on
the merging scale.

\subsection{General Scheme}
\label{sec:basic-steps}

The basic steps are common to all the algorithms and can be summarized as
follows:

\begin{enumerate}

\item Select a process to be studied and choose a scheme to cutoff the
  divergences in the matrix elements, typically using a jet measure.
  Specify the maximum parton multiplicity to be generated by the
  matrix elements (This is currently limited to five or six extra
  partons for computational reasons). Calculate the cross section for
  all the parton multiplicities.

\item Select a parton multiplicity with a probability proportional to
  its integrated cross section. Generate kinematics according to the
  matrix element.

\item Calculate a weight for the event based on the Sudakov form
  factors and the running coupling. Use this weight either to reweight
  the event or as a probability for rejecting the event. If the event
  is rejected generate a new event according to step 2.

\item Find a set of initial conditions for the parton shower and invoke
  the shower. This may include a veto on the emissions from the
  shower.

\end{enumerate}

All the algorithms considered in this paper do steps 1 and 2 in the same
way, but steps 3 and 4 are done using rather different approaches. Each
algorithm has its own way of including the Sudakov form factors and
finding a good set of initial conditions for the shower.

One of the key features that distinguishes the algorithms is the
choice of scales. The algorithms uses different definitions of scales
when determining how to split phase space between the matrix element
and the parton shower and in calculating the Sudakov form
factors. Furthermore, these scale definitions may be different from
what determines the ordering of emissions in the parton showers. We
show later in this paper that the scale choices have significant
consequences for how well the dependence on the merging scale can be
minimized.

The rest of this section describes the details of each algorithm and
their consequences for the physics result. The descriptions of the
algorithms are limited to $e^+e^-$ collisions, but, with a few
modifications and extensions, they have all been used to calculate
results for hadron collisions.

\subsection{CKKW}
\label{sec:theory_ckkw}

The theoretical foundation for CKKW was published in
\cite{Catani:2001cc}, but the main points are repeated here for
completeness. CKKW is focuses on the Durham \kT-algorithm for
clustering jets in $e^+e^-$ \cite{Catani:1991hj}, where the distance
between two partons, $i$ and $j$, is defined as
\begin{eqnarray}
y_{ij} \equiv 2 \min(E_i^2,E_j^2)
(1-cos\theta_{ij})/\ECM^2.\label{eqn:kt}
\end{eqnarray}
The \kT-algorithm is used to construct a parton shower history from
the event generated according to the matrix element. The algorithm
generates a set of clusterings and corresponding scales, which are
later used to calculate the Sudakov form factors and the running
coupling.

Before going through the CKKW algorithm some notation needs to be
introduced. $\Gamma_q$, $\Gamma_g$ and $\Gamma_f$ are the
branching probabilities for $q\rightarrow qg$, $g\rightarrow gg$
and $g\rightarrow f\bar{f}$ respectively. The Sudakov form factors
are then given by
\begin{eqnarray}
\Delta_q(Q_1, Q_2) & = & \exp(-\int_{Q_2}^{Q_1}dq
\Gamma_q(Q_1, q))\\
\Delta_g(Q_1, Q_2) & = & \exp(-\int_{Q_2}^{Q_1}dq
(\Gamma_g(Q_1, q)+\Gamma_f(Q_1, q))).
\end{eqnarray}
They can be interpreted as the probability of a parton with the
production scale $Q_1$ not to have a branching above the scale
$Q_2$. Note the dependency on the production scale in the
branching probabilities. This means that the Sudakov form factors used in
CKKW do not factorize ($\Delta(Q_1, Q_2)\cdot \Delta(Q_2, Q_3) \neq
\Delta(Q_1, Q_3)$). This is also the case for the Sudakov form factors
in the angular ordered shower, where the limits on the integration of
the splitting functions are dependent on the production scale.

The idea of CKKW is to use the full matrix element for the branching
probabilities and analytical Sudakov form factors above the merging
scale, and the parton shower below the merging scale. The full CKKW
algorithm is the following:

\begin{enumerate}

\item Calculate cross sections and generate events according to
step 1 and 2 in section \ref{sec:basic-steps}. \yms\ denotes the
merging scale which is equal to the matrix element cutoff,
defined in terms of the \kT-measure in \eqref{eqn:kt}.
The events are generated using a fixed strong coupling, $\asme$,
and a maximum multiplicity, $N$.

\item Construct a shower history by applying the \kT-algorithm to the
  state from the matrix element. The algorithm is constrained to only
  allow clusterings of partons which are consistent with a possible
  emission from the parton shower. This yields a set of clustering
  values $y_2, ..., y_n$, where $y_2 = 1 > y_3 > ...  > y_n$ and $n$
  denotes the parton multiplicity of the event from the matrix
  element. Use the result from the clustering to determine a set of
  nodes where the partons are merged and the associated scales, $q_i^2
  = y_i \ECM^2$.

\item Calculate a weight for the running coupling given by
  $\prod_{i=3}^n \alpha_s(q_i)/\asme^{n-2}$.

\item For each internal line of type $i$ running between a node with
  scale $q_j$ and $q_k$ apply a factor $\Delta_i(q_j,\qms) /
  \Delta_i(q_k,\qms)$, where $\qms^2 = \yms\ECM^2$. For external
  lines of type $i$ starting from a scale $q_j$ apply the weight
  $\Delta_i(q_j,\qms)$. These weights are the Sudakov form factors and
  they are calculated analytically.

\item Reweight the event with the product of the Sudakov form factors
  in step 4 and the running coupling weight in step 3.

\item Set the starting scale of each parton to the scale associated
  with the node in the shower history where it was produced. Invoke the
  shower and veto any emission which would give a \kT-measure above
  $\yms$.

\end{enumerate}

The original CKKW procedure\cite{Catani:2001cc} contained no special
treatment for highest multiplicity events. This needs to be included
since applying the veto on emissions in the shower down to the merging
scale would prevent events with more than $N$ jets above the merging
scale from being generated. One way to resolve this is to modify the
procedure for highest multiplicity events ($n=N$) and use the scale
$q_n$ instead of $\qms$ in the Sudakov form factors and the vetoed
shower. This is done in \cite{Mrenna:2003if, Schaelicke:2005nv}.

Another issue that needs to be addressed is effects related to the
choice of ordering variable in the shower. The entire theoretical
derivation in the CKKW publication\cite{Catani:2001cc} uses only one way
of defining scales, namely the Durham \kT. While the discrepancy from
using a different ordering variable in the shower may cancel to some
accuracy, this is not explicitly shown or discussed, even though the
shower used in the publication is ordered in virtuality. The
consequences of applying the scheme to a shower not ordered in Durham
\kT\ are therfore somewhat unclear.

The original CKKW publication\cite{Catani:2001cc} claims that this
procedure cancels the dependence on the merging scale to
next-to-leading logarithmic (NLL) accuracy.  However, this claim
assumes that the formalism used, including Sudakov form factors, jet
rates and generating functions, is valid at NLL, but this proof has
never been published.\footnote{The only reference leads to reference 22 in
  \cite{Catani:1991hj}, which is marked ``in preparation'', and it has
  been confirmed by one of the authors of the article that it was
  never completed.}

\subsection{CKKW-L}
\label{sec:theory_ckkw-l}

The CKKW-L algorithm goes through the same basic steps as CKKW, but has
a different way of calculating the Sudakov form factors and implementing
the veto in the shower. In the CKKW-L scheme, a full cascade history
with intermediate states is constructed. What is done is basically to
run the cascade backwards and answer the question ``how could the shower
have generated this state?''. This means that the ordering scale in the
shower is used when clustering the partons. As in the CKKW case, only
physically allowed clusterings are considered. However, contrary to CKKW
where always the smallest scale is chosen for each clustering, all
possible ordered shower histories are considered, and one is chosen
according to a probability proportional to the product of the relevant
branching probabilities. The chosen shower history is then used to
calculate the Sudakov form factors and the running coupling.

To be able to use this scheme, the parton shower needs to have well
defined intermediate states and it is also required that the Sudakov
form factors factorize ($\Delta(Q_1, Q_2)\cdot \Delta(Q_2, Q_3) =
\Delta(Q_1, Q_3)$). This is achieved if the Sudakov form factors only
depend on the kinematics of the intermediate state rather than the
production scale. This is for example the case in the dipole shower used
in \ariadne \cite{ARIADNE92} and the \pT-ordered shower in \pythia
\cite{Sjostrand:2006za}, but not for the angular ordered shower in
\herwig \cite{Corcella:2000bw}.

The effects of the running coupling are taken into account by
reweighting with the same \as\ as in the shower with the constructed
scales as input. The Sudakov form factors are introduced by using the
shower to generate single emissions from the constructed states
starting from the constructed scales and rejecting the event if the
emission is above the next constructed scale. This is known as the
Sudakov veto algorithm and it is equivalent to accepting events with a
probability equal to the Sudakov form factor, since by definition the no
emission probability of the shower is equal to the Sudakov form factor.

There are several advantages to this approach. One is that it makes
sure that the Sudakov form factors above and below the merging scale
match exactly and another is that any corrections introduced in the
shower is also included in the Sudakov form factor. This is
particularly useful if the splitting functions have been reweighted
with matrix elements, since this makes it possible to completely cancel
the merging scale dependence for first order correction (shown explicitly
below). We also expect that the cancellation of the first order
correction will lead to a smaller merging scale dependence for
higher order processes even though the complete cancellation no longer
holds.

We use a slightly different notation in this section to emphasize the
difference in the Sudakov form factors with respect to the ones in CKKW.
$\Delta_{S_n}(\rho_1, \rho_2)$ denotes a Sudakov form factor for an
$n$-particle state giving the probability that there is no emission with
a shower scale, $\rho$, between $\rho_1$ and $\rho_2$. The merging scale
uses another notation, \Qms, to emphasize that this scale does not need
to be defined in terms of the emission scale in the shower. In fact, one
could in principle use any partonic scale definition for the merging
scale. These are the steps in the CKKW-L algorithm.

\begin{enumerate}

\item Calculate cross sections and generate events according to step 1
  and 2 in section \ref{sec:basic-steps}. \Qms\ denotes the merging
  scale which is equal to the matrix element cutoff and may be
  defined using any choice of scale. The events are generated using a
  fixed strong coupling, $\asme$, and a maximum parton multiplicity,
  $N$.

\item Construct a full cascade history by considering all possible
  ordered histories and selecting one randomly with a probability
  proportional to the product of the branching probabilities. If no
  ordered histories can be constructed, unordered ones are considered.
  This results in a set of intermediate states ($S_2, S_3, ... S_n$)
  and scales ($\rho_2 = \rmax, \rho_3, ..., \rho_n$). $S_2$ denotes
  here denotes the constructed $2\rightarrow 2$ process and $S_n$ is
  the state given by the matrix element. $n$ is the parton
  multiplicity in the event, $\rmax$ is the maximum scale of the
  process and $\rho_i$ is the constructed scale where the state
  $S_{i-1}$ emits a parton to produce the state $S_i$.

\item Reweight the events with $\prod_{i=3}^n \alpha_s(\rho_i) /
  \asme^{n-2}$.

\item For each state $S_i$ (except $S_n$), generate an emission with
  $\rho_i$ as starting scale and if this emission occurred at a scale
  larger than $\rho_{i+1}$ reject the event. This is equivalent to
  reweighting with a factor $\prod_{i=2}^{n-1} \Delta_{S_i}(\rho_i,
  \rho_{i+1})$.

\item For the last step there are two cases.

\begin{itemize}

\item If the event does not have the highest multiplicity $n <
N$, generate an emission from the state $S_n$ with $\rho_n$ as
starting scale. If the emission is above the merging scale
$\Qms$, reject the event. Otherwise accept the event and
continue the cascade.

\item If the event has the highest possible multiplicity $n = N$,
accept the event and start the cascade from the state $S_n$ with
the scale $\rho_n$.

\end{itemize}

\end{enumerate}

The algorithm introduces all the factors that would have been present
if the event had been generated by the parton shower, except the
branching probability which is taken from the matrix element. To show
how this comes about, a derivation of the parton multiplicity cross
sections for a first order matrix element is shown below. The explicit
reweighting of \as\ is not shown, but is straight forward to include.

Let $\rho_0$ denote the parton shower cutoff and $\Gamma_{S_n}(\rho)$
denote the probability that a state $S_n$ branches at scale $\rho$.
The definition of the Sudakov from factor in this case is
$\Delta_{S_n}(\rho_1, \rho_2) \equiv \exp( - \int_{\rho_2}^{\rho_1}
d\rho \, \Gamma_{S_n}(\rho))$. The exclusive parton multiplicity cross
sections generated by the standard parton shower can be written as
\begin{eqnarray}
\sps_2(\rmax, \rho_0) & = & \sigma_0 \,
\Delta_{S_2}(\rmax, \rho_0)\label{eq:PS2}\\
\sps_3(\rmax, \rho_0) & = & \sigma_0
\int_{\rho_0}^{\rmax} d\rho \, \Delta_{S_2}(\rmax, \rho)
\, \Gamma_{S_2}(\rho) \, \Delta_{S_3}(\rho, \rho_0)\\
\sps_4(\rmax, \rho_0) & = & \sigma_0
\int_{\rho_0}^{\rmax} d\rho \, \Delta_{S_2}(\rmax, \rho)
\, \Gamma_{S_2}(\rho) \times \nonumber\\
& & \times \int_{\rho_0}^{\rho} d\rho' \,
\Delta_{S_3}(\rho, \rho') \, \Gamma_{S_3}(\rho') \,
\Delta_{S_4}(\rho', \rho_0).
\end{eqnarray}
This notation can be generalized to include all higher multiplicity
cross sections. Let $\evol{3}{n}(\rho, \rho_0)$ denote the
probability of the three-parton state to evolve into an $n$ parton
state between the two given scales. This means that a general $n\ge3$
parton cross section can be written as
\begin{eqnarray}
\sps_n(\rmax, \rho_0) & = & \sigma_0
\int_{\rho_0}^{\rmax} d\rho \, \Delta_{S_2}(\rmax, \rho)
\, \Gamma_{S_2}(\rho) \evol{3}{n}(\rho, \rho_0),\label{eq:PSn}
\end{eqnarray}
where
\begin{eqnarray}
\evol{3}{3}(\rho, \rho_0) & = & \Delta_{S_3}(\rho, \rho_0).\\
\evol{3}{4}(\rho, \rho_0) & = & \int_{\rho_0}^{\rho}d\rho'\,
\Delta_{S_3}(\rho, \rho')\, \Gamma_{S_3}(\rho')\,
\Delta_{S_4}(\rho', \rho_0)\\
\vdots\nonumber
\end{eqnarray}

The merging scale may be defined using a different way of mapping the
phase space (denoted $Q$) as compared to the scale used in the shower.
The value of the scale used to define the merging scale and the scale of
the shower can be related if the other variables that determine the
shower emission is included. Let $\vec{x}$ represent the $k$ additional
variables used in the shower, which may include energy fractions and
rotation angles. The branching probability can be written in a way that
it includes the dependence on these variables as long as the following
is true.
\begin{eqnarray}
\Gamma(\rho) = \int d^k\vec{x}\, \Gamma(\rho, \vec{x})
\end{eqnarray}

The alternative mapping of phase space is described by the function
$Q_{S_n}(\rho, \vec{x})$, which denotes the value of the merging scale
measure for a given shower emission. The lowest order matrix element
cross sections are equal to
\begin{eqnarray}
\sme_2(\rmax, \Qms) & = & \sigma_0\\
\sme_3(\rmax, \Qms) & = & \sigma_0
\int_{\rho_0}^{\rmax} d\rho \int d^k\vec{x} \, 
\Theta(Q_{S_2}(\rho, \vec{x}) -
\Qms) \, \gme_{S_2}(\rho, \vec{x}). \label{eqn:me3jet}
\end{eqnarray}
$\gme_{S_2}$ is the branching probability in the matrix element and
$\Theta$ is the standard Heaviside function. Note that the equations
above only hold if the matrix element merging scale is above the
shower cutoff everywhere in phase space, but this can be resolved by
discarding all events from the matrix element which are below the shower
cutoff.

The next step is to apply the algorithm to calculate the jet rates at
$\rho_0$ for the merged matrix element and parton shower. The two-jet
state is already the lowest order process, which means no cascade
history is constructed and only the final Sudakov veto down to the
merging scale enters. The parton multiplicity cross sections become
equal to that of the pure shower minus the cross section for events with
the first emission above \Qms. The two-jet matrix-element contributions
to the cross sections are equal to
\begin{eqnarray}
\sigma_2(\rmax, \rho_0) & = & \sigma_0 \, 
\Delta_{S_2}(\rmax, \rho_0)\\
\sigma_n(\rmax, \rho_0) & = & \sigma_0
\int_{\rho_0}^{\rmax} d\rho \int d^k\vec{x} \, \Theta(\Qms -
Q_{S_2}(\rho, \vec{x})) \, \Delta_{S_2}(\rmax, \rho)\times
\nonumber\\
& & \times \Gamma_{S_2}(\rho, \vec{x}) \evol{3}{n}(\rho, \rho_0).
\end{eqnarray}

When the algorithm is applied to a three-jet event, an emission scale,
$\rho_3$, is constructed. Emissions are generated from the two-jet
state and events discarded according to the Sudakov veto algorithm,
which is equivalent to introducing a weight $\Delta_{S_2}(\rmax,
\rho_3)$. The cascade is then started from the scale $\rho_3$ and,
assuming the three-jet configuration is the highest multiplicity, no
additional Sudakov suppression is included.  Using the matrix element
cross section from equation (\ref{eqn:me3jet}) and, assuming that the
constructed scale $\rho_3$ is equal to the scale used in the matrix
element, results in the following contributions to the parton
multiplicity cross sections.
\begin{eqnarray}
\sigma_n(\rmax, \rho_0) & = & \sme_3(\rmax, \Qms)
\, \Delta_{S_2}(\rmax, \rho_3)\, \evol{3}{n}(\rho_3, \rho_0)
= \nonumber\\
& = & \sigma_0 
\int_{\rho_0}^{\rmax} d\rho \int d^k\vec{x}\, \Theta(Q_{S_2}(\rho,
\vec{x}) - \Qms) \, \gme_{S_2}(\rho, \vec{x})
\times\nonumber\\
& & \times \Delta_{S_2}(\rmax, \rho)\, \evol{3}{n}(\rho, \rho_0)
\end{eqnarray}

The following cross sections are the result from adding the
contributions from the two- and three-jet processes.
\begin{eqnarray}
\smeps_2(\rmax, \rho_0) & = & \sigma_0 \,
\Delta_{S_2}(\rmax, \rho_0)\\
\smeps_n(\rmax, \rho_0) & = & \sigma_0 \,
\int_{\rho_0}^{\rmax} d\rho \int d^k\vec{x} \, 
[\gme_{S_2}(\rho, \vec{x}) \, 
\Theta(Q_{S_2}(\rho, \vec{x}) - \Qms) +\nonumber\\
& & + \gps_{S_2}(\rho, \vec{x})
\Theta(\Qms - Q_{S_2}(\rho, \vec{x}))] \,
\Delta_{S_2}(\rmax, \rho)
\, \evol{3}{n}(\rho, \rho_0)
\end{eqnarray}

From the equations above one can see that the only dependence on
the merging scale \Qms\ is in the integration over the branching
probabilities. This means that the merging scale dependence for
this process cancels to the accuracy of which the shower
generates the first emission. In fact, for the process studied in
this paper, many of the parton shower implementations reweight
the splitting function for the first emission with the matrix
element, in which case the merging scale dependence cancels
completely.

\subsection{Pseudo-Shower}
\label{sec:pseudo-shower}

The Pseudo-Shower algorithm is similar in spirit to CKKW-L, but it uses
partonic jet observables for the kinematics instead of individual
partons. Where CKKW-L runs the cascade one emission at a time from a
given parton state to calculate no-emission probabilities, in the
Pseudo-Shower approach a full parton cascade is evolved, and the
resulting partons are clustered back to jets. Any standard clustering
algorithm, where in each step the pair of particles which are closest
together are clustered, can be used. The same distance measure is
used to define the merging scale as the one used to construct a
shower history for the matrix-element state and the fully showered
states. The full algorithm is defined as follows.

\begin{enumerate}

\item Choose a jet clustering scheme to be used in the algorithm and a
  merging scale, \dms. Set the matrix element cutoff equal to the
  merging scale, calculate cross sections and generate events
  according to step 1 and 2 in section \ref{sec:basic-steps}. The events
  are generated using a fixed strong coupling, $\asme$, and a maximum
  multiplicity, $N$.

\item\label{step:ps-clus} Cluster the partons from the matrix
  element, using the selected jet scheme, until a $2\rightarrow 2$
  state is reached. As in CKKW, only clusterings corresponding to
  physically allowed splittings are considered.  The result of the
  clustering is interpreted as a shower history with a set of states
  ($S_2, \ldots S_n$) and a set of scales ($\tilde{d}_2=\dmax,
  \tilde{d}_3,\ldots , \tilde{d}_{n}$), where $n$ is the parton
  multiplicity of the event.

\item\label{step:ps-sudakov} For each state $S_i$, except $S_n$,
  perform a full shower vetoing any emission with a jet measure
  greater than the corresponding scale in the shower history, $d >
  \tilde{d}_i$. Calculate a set of clustering scales, $d_j$, by
  clustering the partons from the shower using the same algorithm as
  in step \ref{step:ps-clus}, but for practical reasons also allow
  clusterings corresponding to non-physical splittings.  Reject the
  event if $\sqrt{d_{i+1}} > \sqrt{\tilde{d}_{i+1}} + \delta$ (the
  $\delta$ is a fudge factor to be discussed below).

\item For the final state $S_n$ the shower is invoked, vetoing
  emissions with $d > \tilde{d}_n$. If the event is not a maximum
  multiplicity event ($n<N$), the partons are clustered and the event
  is rejected if the scale from the clustering is above the merging
  scale, $\sqrt{d_{n+1}} > \sqrt{\dms}+\delta$.  For $N=n$ the event is
  accepted except if $\sqrt{d_{n+1}}>\sqrt{\tilde{d}_{n}}+\delta$.

\end{enumerate}

Although it was not stated in the text in \cite{Mrenna:2003if}, the
implementation did include a reweighting with a running strong coupling
$\prod_{i=3}^n \as(\tilde{d}_i) / \asme^{n-2}$. The same reweighting is
also included here.

To see the similarity with CKKW-L, consider using the Pseudo-Shower
using a parton shower with an ordering variable equal to the distance
scale in the clustering algorithm used, and with well-defined
intermediate states. In the strongly ordered limit, the clustering
algorithm would then exactly reproduce the intermediate states and
branching scales in the shower. This means that evolving a full shower
from the state $S_i$, starting from a scale $\tilde{d}_i$, clustering
to find a $d_{i+1}$ and rejecting the event if
$d_{i+1}>\tilde{d}_{i+1}$ would exactly correspond to the Sudakov form
factor $\Delta_{S_i}(\tilde{d}_i,\tilde{d}_{i+1})$, and the merging
scale dependence for first order matrix element corrections would cancel
in the same way as CKKW-L.

In reality, the clustering does not exactly reconstruct the shower
splittings. If subsequent emissions are not clustered in the same way as
the shower emitted them, this can affect the clustering scale of harder
emissions. This means that the scale of the matrix element partons
before the shower and the scale that one gets from the jet clustering
after the shower are rarely the same. This is a significant problem,
since the phase space cuts that separate the matrix element emissions
from the partons shower emission is done using two different scales,
which leads to dead regions and double counting of emissions and it also
affects the calculation of the Sudakov form factors.

To moderate the effects of having two different scales, the fudge
factor $\delta$ (introduced in \cite{Mrenna:2003if}) was included
whenever comparing a scale from the clustering of the matrix element
state with a scale from the clustering of the fully showered state. In
\cite{Mrenna:2003if} the value $\delta =2$~GeV was used without
motivation, theoretical or otherwise, but supposedly the parameter
needs to be tuned for each choice of process and merging scale to
properly compensate for the mismatch in scales.

\subsection{MLM}
\label{sec:theory_mlm}

The MLM algorithm is similar to the Pseudo-Shower in that it also does
matching with partonic jet observables. The algorithm is much simpler
to implement compared to earlier schemes discussed. In the MLM merging
scheme the event from the matrix element is simply fed into the parton
shower program, the shower is invoked and the final state partons are
clustered into jets. The algorithm then specifies that the matrix
element partons should be matched to the final state partonic jets, and
events are accepted only if all the jets match and the event contains
no extra jets above the merging scale. In this way the Sudakov form
factors are approximated by the probability that there are no
emissions above the merging scale and, at the same time, the parton
shower emissions are approximately constrained to be below the merging
scale. The MLM algorithm is a really convenient way of doing merging
since it requires no modifications to the parton shower program.

Even though MLM has been frequently used, a general version of the
algorithm has never been published and all the published algorithms
assume incoming hadrons. Based on \cite{MLM, Alwall:2007fs,
  Mangano:2006rw}, we present here our interpretation of the necessary
steps needed for applying the MLM scheme to $e^+e^-$ collisions.

The first step in an MLM implementation is to choose a jet definition to
be used for the merging scale and the matrix element cutoff. The
original MLM algorithm used cone jet definitions, although there have
been implementations using the \kT-algorithm, \eg\ the \madevent
implementation in \cite{Alwall:2007fs} and the \herwig implementation in
\cite{Mrenna:2003if}. After specifying a cutoff, events are generated
and the running coupling reweighting is calculated in the same way as in
the CKKW algorithm.

Then the shower is invoked using an appropriate starting scale, which
is defined for each implementations and process. For $W$-production in
hadron collisions, the scale is set to the transverse mass of the $W$
($\sqrt{m_W^2 + p_{\perp W}^2}$)\cite{Alwall:2007fs}, whereas for the
top production implementation it is not specified
\cite{Mangano:2006rw}. In hadron collision there is some freedom for
choosing the starting scale of the shower, but for $e^+e^-$ we think
that the scale should be set to center of mass energy, to allow the
shower to utilize the full phase space.

After the shower has been invoked the final state partons need to
be clustered into jets and matched to the partons from the matrix
element. The clustering is done with the same algorithms used to
define the merging scale. The partons from the matrix element are
then matched to the clustered jets, in order of decreasing
energy. The measure used to match the partons to jets is some quantity
related to the jet clustering. These are all the steps in the
algorithm.
\begin{enumerate}

\item Select a merging scale, \Qms, and a matrix element cutoff \Qcut,
  such that $\Qcut < \Qms$, where the scales are defined using a jet
  algorithm. Calculate cross sections and generate events according to
  step 1 and 2 in section \ref{sec:basic-steps}. The events
  are generated using a fixed strong coupling, $\asme$, and a maximum
  parton multiplicity, $N$.

\item Cluster the partons from the matrix element using the
  \kT-algorithm and use the clustering scales as in input to \as\ and
  reweight the event.

\item Feed the event into a parton shower using the Les Houches
  interface\cite{Boos:2001cv}, setting the scale to \ECM, and start
  the shower.\label{step:mlm_shower}

\item Cluster the partons to jets using the algorithm from step 1 with
  a clustering scale set to \Qms. Go through the list of partons, in
  order of decreasing energy, and match them to the clustered
  jets. This is done by finding the jet with the smallest distance to
  the parton defined using some measure based on the jet clustering
  scheme\footnote{This cannot be exactly the same distance measure as
    in the jet algorithm for reasons to be discussed in section
    \ref{sec:results}}. If not all the partons match or there are
  extra jets, reject the event.

  For the highest multiplicity events either use a higher clustering
  scale and more relaxed matching criteria or allow extra jets that
  are softer than the matched jets.

\end{enumerate}

There are several aspects of this algorithm that needs further
explaining. The reason for having a cutoff below the merging scale is
that events slightly below the merging scale can end up above after the
shower. This leaves an arbitrary choice of matrix element cutoff, but
this can be resolved if soft and collinear particles have a vanishing
probability to generate an independent jet, which means that the result
converge when the matrix element cutoff is lowered. This is a way of
getting around the problem that occurred in the Pseudo-Shower algorithm,
namely that the cuts on the matrix element state and on the partonic
jets are not equivalent. However, the convergence needs to be verified
for each implementation and process.

One other aspect that needs further scrutiny is what happens inside
the parton shower program. The main danger is that the program may be
given a state with one or more relatively soft parton and a rather high
starting scale, which means that the shower often ends up emitting
harder partons than the ones already present, leading to an unordered
shower. This breaks the strong ordering approximation, which is
fundamental to all parton showers, and the end result is heavily
dependent on how unordered emissions are handled.

To derive some of the properties of MLM, let us assume (as we did in
the Pseudo-Shower case) that the jet clustering is a perfect inverse
of the shower and that the shower has well defined intermediate
states. These assumptions are a bit crude considering the way MLM is
used in current implementations, but it allows for the possibility to
do analytical calculations and it should give some idea of what to
expect from the algorithm. Under these assumptions parton multiplicity
cross sections, including the first order matrix element corrections,
can be calculated.

The calculations are performed using the same notation as in the
section \ref{sec:theory_ckkw-l}. The merging scale can be defined
in terms of the scale in the shower (\rms) and starting the scale
from the center of mass energy is equivalent to using the maximum
scale (\rmax). The two-jet matrix element contribution to the parton
multiplicity cross sections becomes the same as in CKKW-L.
\begin{eqnarray}
\sigma_2(\rmax, \rho_0) & = & \sigma_0 \cdot
\Delta_{S_2}(\rmax, \rho_0)\\
\sigma_n(\rmax, \rho_0) & = & \sigma_0 \cdot
\int_{\rho_0}^{\rms} d\rho \, \Delta_{S_2}(\rmax,
\rho)
\, \Gamma_{S_2}(\rho) \evol{3}{n}(\rho, \rho_0)
\end{eqnarray}
The contribution from the three-jet is different however. The reason
is that there is no Sudakov form factor from a two-particle state
included since no shower history was considered. The contribution to
the cross sections is the following.
\begin{eqnarray}
\sigma_n(\rmax, \rho_0) & = & \sigma_0 \cdot
\int_{\rms}^{\rmax} d\rho \, \gme_{S_2}(\rho) \,
\Delta_{S_3}(\rmax, \rho)\, \evol{3}{n}(\rho, \rho_0)
\end{eqnarray}
The sum of the two contributions become the following. 
\begin{eqnarray}
\sigma_2(\rmax, \rho_0) & = & \sigma_0 \cdot
\Delta_{S_2}(\rmax, \rho_0)\label{eq:MLM_MEPS2}\\
\sigma_n(\rmax, \rho_0) & = & \sigma_0 \cdot
\int_{\rho_0}^{\rmax} d\rho \, \left[\gme_{S_2}(\rho)\,
\Delta_{S_3}(\rmax, \rho)
\Theta(\rho - \rms) +\right. \nonumber\\
& & \left.+\gps_{S_2}(\rho)
\, \Delta_{S_2}(\rmax, \rho)
\Theta(\rms - \rho)\right] \evol{3}{n}(\rho, \rho_0)
\label{eq:MLM_MEPSn}
\end{eqnarray}
Comparing equation (\ref{eq:MLM_MEPS2}) to (\ref{eq:PS2}) one can see
that the two-parton cross section becomes the correct one. Note that
this would not have be the case if a lower starting scale was chosen
for the shower. The higher multiplicity cross sections contain
complications, which can be seen by comparing equation
(\ref{eq:MLM_MEPSn}) to (\ref{eq:PSn}). The problem is that MLM does
not include the Sudakov form factor from the $S_2$ state, which means
that there will be an additional dependence on the merging scale as a
result of the difference in the Sudakov form factors. The factor
$\Delta_{S_3}(\rmax, \rho)$ is where the explicitly unordered shower
occurs and the results are therefore largely dependent on the parton
shower implementation.

Consider, for illustration, using the \kT-ordered shower of \pythia on
a three-parton state. Here the maximum transverse momentum of an
emission is given by half the largest of the $qg$ and $g\bar{q}$
invariant masses, which for a soft gluon can be very small. Hence, the
Sudakov form factor between \ECM\ and this transverse momentum
would be absent, resulting in a large dependence on the merging scale.

The actual MLM implementations contain several other
complications. The jet clustering used is not the inverse of the
shower and most implementations use parton showers that do not
have well defined intermediate states. However, none of these
aspects can resolve the problem that the Sudakov form factor is
generated using a three-particle state instead of a two-particle
state. The error caused by using the wrong Sudakov form factor is
inherent in any MLM implementation.

\section{Results}
\label{sec:results}

Each of the algorithms described above have been tested for the first
order matrix element correction to $e^+e^- \rightarrow q \bar{q}$ at
the $Z^0$ pole. As explained in the introduction, this matrix element
correction can also be included by a simple reweighting of the first
(or hardest) splitting in a parton cascade, thus providing us with the
``correct'' answer for comparison. In this way we can check whether
the merging algorithms actually meet their goals of a clean cut
between matrix element and parton shower phase space and a small
dependence on the merging scale. Only if they do achieve these goals
on this simple case, can we believe that they are likely to achieve
their goals when generalized to higher order matrix elements and more
complicated processes.

To have a fair comparison, we have in all cases generated matrix
element events using a cut in the Durham \kT-algorithm distance
measure, $y$, as defined in \eqref{eqn:kt}. Also the merging scale
is defined using this distance measure. The only exception is for
the Pseudo-Shower algorithm, where a slightly different scale is used
as explained below in section \ref{sec:res:pseudo}. The matrix element
events were generated with \madevent (v 4.1.31) \cite{Alwall:2007st}.

To check the merging scale dependence we look at the distribution which
should be the most sensitive, namely the $y_3$ scale where the
\kT-algorithm clusters three jets into two, when applied to the final
parton-level events. We also look at two hadron-level event shape
observables, which have been well measured at LEP and corrected to
hadron level. One is the normalized $y_3$ distribution of all final
state particles measured by ALEPH \cite{Heister:2003aj}, which shows how
the dependence on the merging scale is reflected in the hadronic final
state. The other is the normalized charged particle thrust distribution
measured by DELPHI \cite{Abreu:1996na}, which is not directly related to
the merging scale, but is nevertheless very sensitive to the leading
order matrix element correction.

\FIGURE[t]{
  \epsfig{file=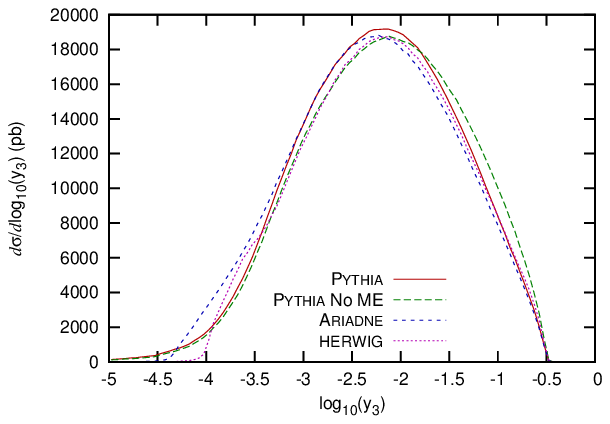,width=10cm}
  \caption{\label{fig:comp_par_y3} The $y_3$ spectra at parton level for
\protect\pythia , \protect\pythia with matrix element reweighting
switched off, \protect\ariadne and \protect\herwig.}
}
\FIGURE[t]{
    \epsfig{file=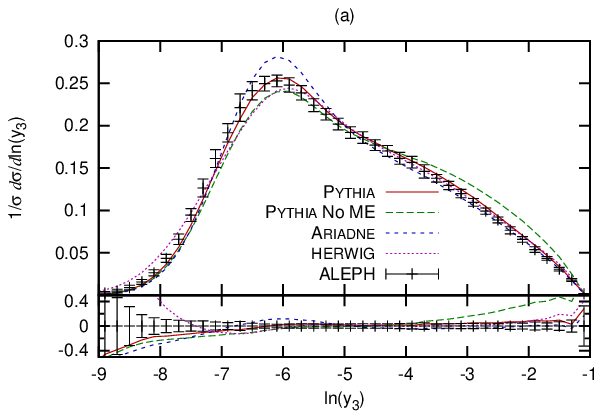,width=0.5\textwidth}%
    \epsfig{file=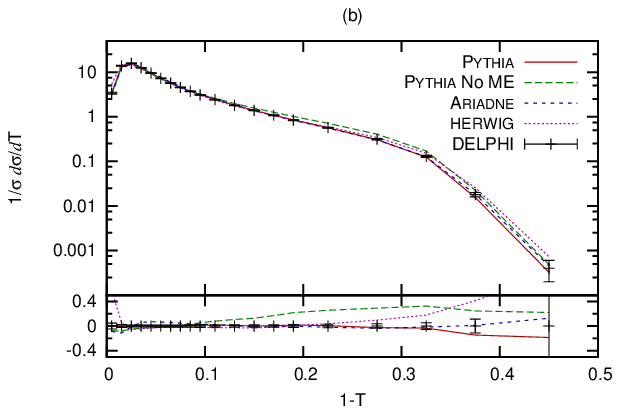,width=0.5\textwidth}
    \caption{\label{fig:comp_had} The $y_3$ spectra for charged and
      neutral particles (a) and the charged particle thrust spectra (b)
      for \protect\pythia , \protect\pythia with matrix element
      reweighting switched off, \protect\ariadne and \protect\herwig
      compared to ALEPH and DELPHI data. The in-sets at the bottom of
      the plots show the relative differences between the Monte Carlo
      results and data, $(\sigma_{\mrm{MC}} - \sigma_{\mrm{Data}}) /
      \sigma_{\mrm{Data}}$.}
}

In figures \ref{fig:comp_par_y3} and \ref{fig:comp_had} we present
these distributions for the four generators \ariadne (v
4.12)\cite{ARIADNE92}, \herwig (v 6.510) \cite{Corcella:2000bw} and
\pythia (v 6.413) \cite{Sjostrand:2006za}, which all are equipped with
simple matrix element reweighting. We see that they all agree fairly
well, which should come as no surprise since they have all been tuned
to fit LEP data.

In figures \ref{fig:comp_par_y3} and \ref{fig:comp_had} we also show
the results from \pythia, with the matrix element reweighting switched
off to give a sense of how large an effects we should expect from the
matrix element corrections. We see that the main effect is that these
distributions are clearly harder than the ones with matrix element
reweighting.

We had planed to also include \sherpa \cite{Gleisberg:2003xi} in this
comparison and use it for the CKKW results. Unfortunately we discovered
inconsistencies\footnote{Both the version 1.0.10 and 1.0.11 of
\protect\sherpa give different results for the cross section depending
on whether weighted or unweighted events are used. Changing between
weighted and unweighted events also gives two different results for the
shape observables in version 1.0.10, neither being consistent
with the results from version 1.0.11.} in the results and therefore
decided not to use \sherpa at all in this study.

\subsection{CKKW-L}
\label{sec:res-ckkwl}

We start by looking at the results from the CKKW-L scheme. Two
implementations have been considered. We have used the original
implementation\cite{Lonnblad:2001iq} in \ariadne and we have also made
an implementation of the first order corrections using the transverse
momentum ordered shower in \pythia \cite{Sjostrand:2006za,
  Sjostrand:2004ef}.

In \ariadne the parton evolution is modeled by a dipole cascade
\cite{Gustafson:1986db,Gustafson:1988rq}. Unlike most other parton
showers, the dipole cascade is based around $2\rightarrow 3$ partonic
splittings rather than $1 \rightarrow 2$. This model automatically
includes the coherence effects from emitting gluons from
colour-neighboring partons, which means that explicit angular ordering
is not necessary. It also means that the first order matrix element
$e^+e^- \rightarrow qg\bar{q}$ is already present in the cascade by
construction. The ordering variable used in the cascade is a
Lorentz-invariant transverse momentum measure
\begin{eqnarray}
p_\perp^2 = \frac{s_{12}s_{23}}{s_{123}},
\end{eqnarray}
where $s_{ij}$ are the invariant masses of the partons and index 2
indicates the emitted parton. This measure is then also used in the
construction of the shower history in a procedure similar to the
\diclus jet clustering algorithm\cite{Lonnblad:1993qd}, but including
only physically allowed clusterings.\footnote{For higher
  multiplicities, all possible shower histories are
  considered. However, for this simple case there is only one unique
  history}

We have also implemented the CKKW-L scheme using the
transverse-momentum ordered shower\cite{Sjostrand:2004ef} in
\pythia.\footnote{Note that this is not a complete implementation,
  since only the leading order matrix element correction to $e^+e^-\to
  q\bar{q}$ is considered.} This cascade exhibits the main features
required by the CKKW-L scheme, namely that it is ordered in transverse
momentum and that is has well-defined on-shell intermediate
states. The ordering variable (also called \pT) is defined in the
following way:
\begin{eqnarray}
\pT^2 = z(1-z)Q^2
\label{eq:pytpt}
\end{eqnarray}
$Q^2$ is the invariant mass of the produced parton and radiating
parton and $z$ is the energy fraction of the radiating parton. The
ordering of the shower is done in a way which incorporates coherence
effects without requiring explicit angular ordering.

\FIGURE[t]{
  \epsfig{file=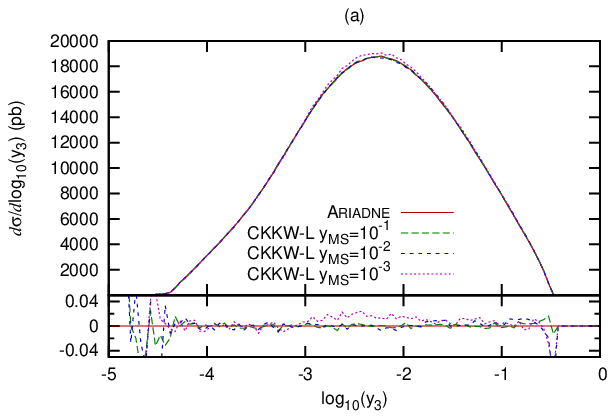,width=0.5\textwidth}%
  \epsfig{file=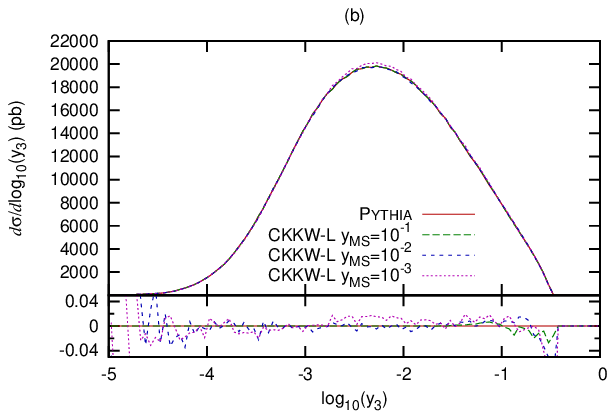,width=0.5\textwidth}
  \caption{\label{fig:ckkwl-par} (a) The $y_3$ spectra at parton level
    for \protect\ariadne (the first order matrix element is included
    by construction) and for the \protect\ariadne implementation of
    CKKW-L corrections with different merging scales. (b) is the same
    for our \protect\pythia implementation of CKKW-L corrections with
    different merging scales. The in-sets at the bottom of the plots
    show the relative differences between the CKKW-L results and
    the default shower, $(\sigma_{\mrm{CKKW-L}} - \sigma_{\mrm{Shower}})
    / \sigma_{\mrm{Shower}}$.}
}

There is one significant difference in the \pythia implementation,
namely that the constructed history is no longer unique. In \pythia
there are two possible ways for the gluon to be emitted (from the quark
or the anti-quark), which means that there are two possible histories to
be considered. One history is selected with a probability proportional
to its branching probability.

In figure \ref{fig:ckkwl-par} we show the parton-level $y_3$
distribution for \ariadne and transverse momentum ordered \pythia
shower with matrix element reweighting.\footnote{We have used \pythia
  v 6.413, amended with a fix approved by the authors to avoid a bug
  in the built-in matrix element reweighting. This bugfix has been
  included v 6.414.} Both programs are compared to their respective
CKKW-L implementations with different merging scales, and the figure
shows that the cancellation is almost complete. There are some small
discrepancy for the lowest cutoff of the order of 2\%. This is because
\madevent generated the matrix element with massless u, d, s and c
quarks, where as in \ariadne and \pythia they are given a small mass,
which causes a slight deviation in the emission probability. This
discrepancy can be removed by setting the masses in \ariadne and
\pythia equal to the masses in \madevent. No deviations, however, are
visible below the merging scale.

\FIGURE[t]{
    \epsfig{file=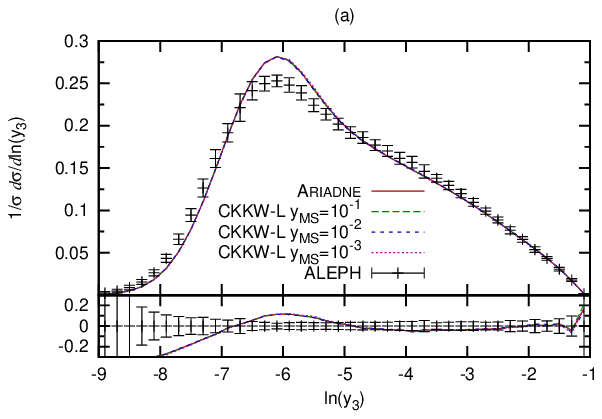,width=0.5\textwidth}%
    \epsfig{file=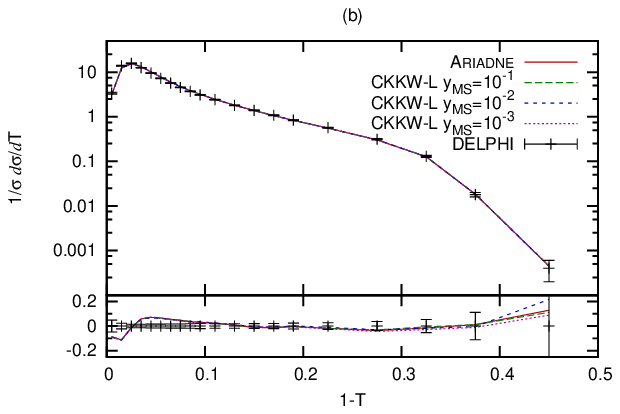,width=0.5\textwidth}
    \epsfig{file=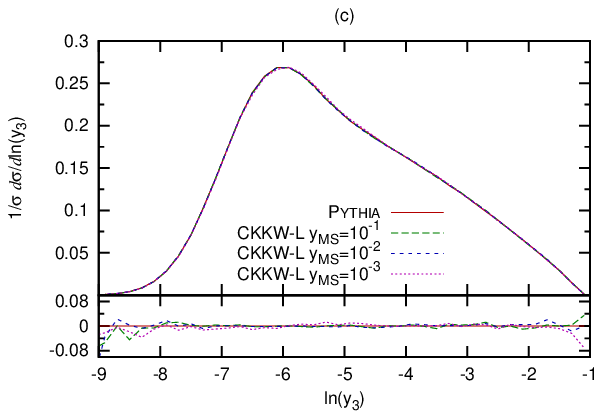,width=0.5\textwidth}%
    \epsfig{file=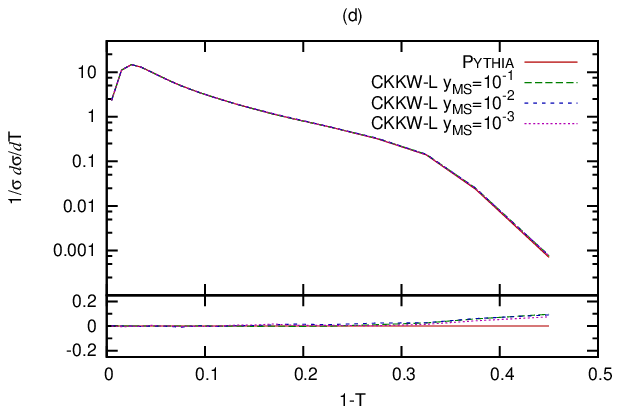,width=0.5\textwidth}
    \caption{\label{fig:ckkwl-had} Charged plus neutral particle $y_3$
and charged particle thrust for the \protect\ariadne and \protect\pythia
implementation of CKKW-L corrections with different merging scales.
Figure (a) includes a comparison to ALEPH data and (b) a comparison to
DELPHI data, where figure (c) and (d) only compare to the default
\protect\pythia, since the \pT-ordered shower has not yet been tuned to
data.}
}

For completeness we show in figure \ref{fig:ckkwl-had} the comparison
for the hadron-level observables $y_3$ and thrust for \ariadne and
\pT-ordered \pythia respectively. As expected from the parton-level
results, there is no serious dependence on the merging
scale.\footnote{The \protect\pythia \pT-ordered shower has not been
  properly tuned to LEP data, and we therefore do not compare it
  directly to data. However, comparing with figures
  \ref{fig:ckkwl-had}a and b, it is clear that the variations due to
  the merging scale are well within the experimental errors.}

\subsection{CKKW}

\FIGURE[t]{
  \epsfig{file=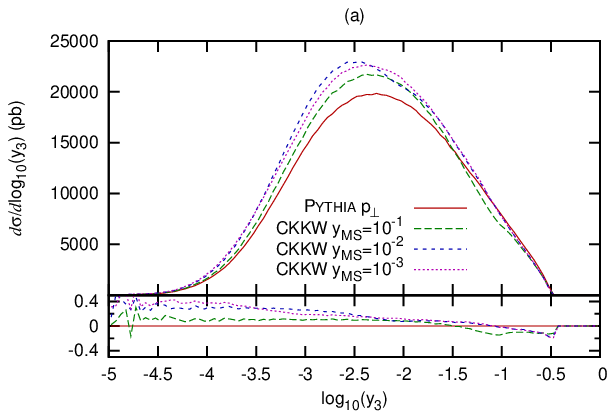,width=0.5\textwidth}%
  \epsfig{file=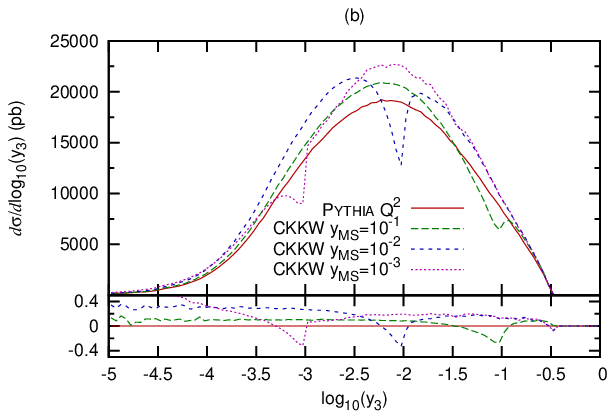,width=0.5\textwidth}
  \caption{\label{fig:ckkw_py_par_y3} The $y_3$ spectra at parton
    level for \protect\pythia with \pT-ordered (a) and virtuality
    ordered (b) shower (and with the first order matrix element
    reweighting included) and for our corresponding \protect\pythia
    implementations of CKKW corrections with different merging
    scales.}
}

The CKKW algorithm has been implemented using \pythia, with both the
\pT-ordered and the virtuality ordered shower. The implementations are
done according to the scheme described in section
\ref{sec:theory_ckkw}.  The only difference is that none of the
implementations use the Durham \kT\ as ordering variables, therefore
setting the starting scale is done differently. The scale of the quark
and anti-quark is set to $\ECM$ and the scale of the gluon is set to
the \pT\ of the reconstructed splitting for the \pT-ordered shower and
the virtuality and angle\footnote{Besides having the virtuality as
  ordering variable, \pythia also imposes a veto on emission angles to
  ensure angular ordering.} of the splitting for the virtuality
ordered shower.  For the latter, the scheme is essentially equivalent
to what is implemented in \sherpa. The results are shown in figure
\ref{fig:ckkw_py_par_y3}.

The results from the \pT\ ordered shower show a smooth transition
between the regions above and below the cutoff. This is to be expected
since Durham \kT\ is approximately equal to the \pT\ in the shower, which
means that the entire procedure becomes similar to CKKW-L with the
exception that the Sudakov form factors are calculated according to an
analytical approximation. The CKKW results show a slightly higher cross
section than standard \pythia, which is attributed to the fact that in
the analytical Sudakov form factors the approximate splitting functions
are integrated over parts of phase space where they are negative. This
means that Sudakov form factors cannot be interpreted as no-emission
probabilities in the same way as in the shower and the end result is a
slightly smaller suppression.

Figure \ref{fig:ckkw_py_par_y3}b shows the results from the virtuality
ordered shower and there are clear problems. Each CKKW curve has a dip
right at the merging scale and for higher values of $y_3$ they are
significantly above the default results from \pythia. One source of
problems is that the shower can generate unordered emissions, since an
emission which has a high vitruality can have a low $k_\perp$. This also
changes which Sudakov form factor is applied to each region of phase
space. A derivation of the problems that this can lead to for the lowest
order process is presented in appendix \ref{sec:appendix}.

\FIGURE[t]{
    \epsfig{file=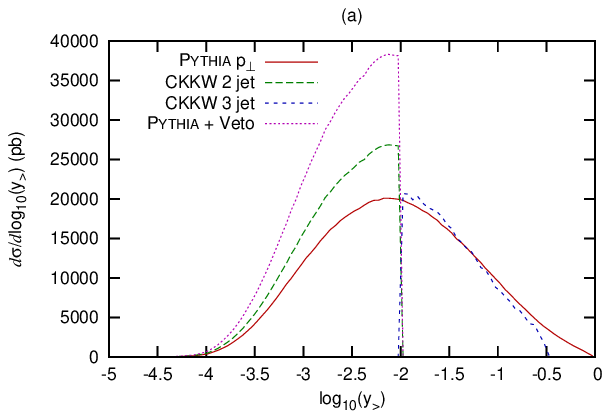,width=0.5\textwidth}%
    \epsfig{file=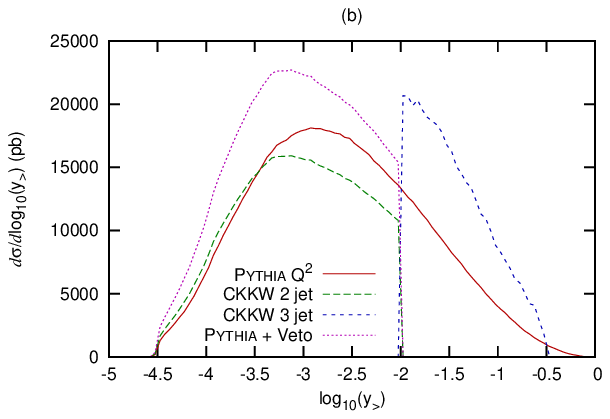,width=0.5\textwidth}
    \caption{\label{fig:ckkw_py_max_y3} The maximum Durham \kT-value of any
    emission in the cascade for
    \protect\pythia with \pT-ordered shower (and with the first order
    matrix element reweighting included) and for the two- and three-jet
    components in our \protect\pythia implementation of CKKW corrections
    with a merging scale $\yms = 10^{-2}$. An extra curve is included to
    show the two-jet contribution without the suppression from the analytic
    Sudakov form factor.}
}

To further scrutinize the causes of the problems in the virtuality
ordered CKKW implementation, we have studied the variable used for the
shower veto which is the $k_\perp$-value of the individual emission.
Figure \ref{fig:ckkw_py_max_y3} shows the maximum Durham \kT-value for
the emissions in the shower. The results have been split into two and
three-jet components of the CKKW implementation using \pythia \pT\ and
virtuality ordered shower.

The two-jet events in CKKW is produced by invoking the shower vetoing
emissions above the merging scale and reweighting with an analytical
Sudakov form factor. These two steps can be considered separately, which
is illustrated in figure \ref{fig:ckkw_py_max_y3} by having a curve
that only includes the veto and not the reweighting.

For the shower ordered in transverse momentum the three-jet
contribution match the curve from running default \pythia fairly well
above the merging scale. This is expected since the ordering in the
transverse momentum ordered shower is fairly close to using Durham
\kT-ordering.  The two-jet contribution is similar in shape but is
higher in cross section, which can be attributed to a smaller
suppression from the analytical Sudakov form factors as compared to
the form factors used in the shower. There is a discontinuity at the
merging scale, but it is smoothed out by the shower causing the figure
\ref{fig:ckkw_py_par_y3}a to appear somewhat smooth.

With the virtuality ordered shower the results are quite different,
which is caused by using two rather different scales. The problem is
that emissions modifies the phase space for subsequent emissions, and by
emitting the partons with highest virtuality first, the hardest
emissions in \kT\ are no longer allowed. When studying the hardest
emission according to Durham \kT, this results in a shift to smaller
values. This is why the \pythia curve is significantly below the
three-jet CKKW curve in figure \ref{fig:ckkw_py_max_y3}b. The sum of the
two and three-jet contributions in figure \ref{fig:ckkw_py_max_y3}b has
two peaks with a clear dip in between. When the rest of the emissions
are included this structure is somewhat smothed out, but the two peaks
with a dip in between are clearly visible in figure
\ref{fig:ckkw_py_par_y3}b.

It is clear that one has to be a bit careful regarding the choice of
shower and scales in CKKW. One CKKW implementation
based on \herwig was published in \cite{Mrenna:2003if} and the results
were consistent only after a significant amount of tuning of scale
parameters. A similar procedure could probably be used to make our
results more consistent, but it would add extra somewhat arbitrary
parameters to the model. The problem with the ordering of the
emissions also leads to a different colour structure in the events, which
was pointed out in \cite{Nason:2004rx}.

\FIGURE[t]{
    \epsfig{file=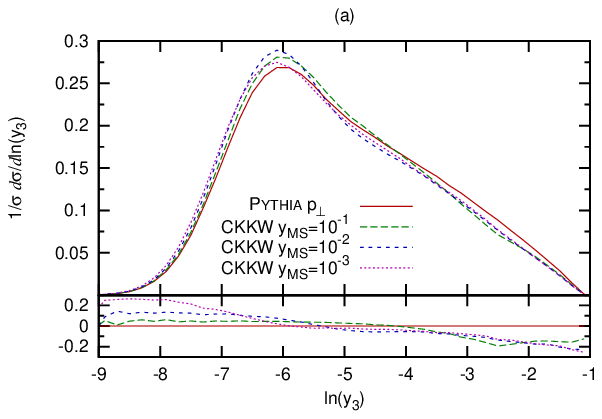,width=0.5\textwidth}%
    \epsfig{file=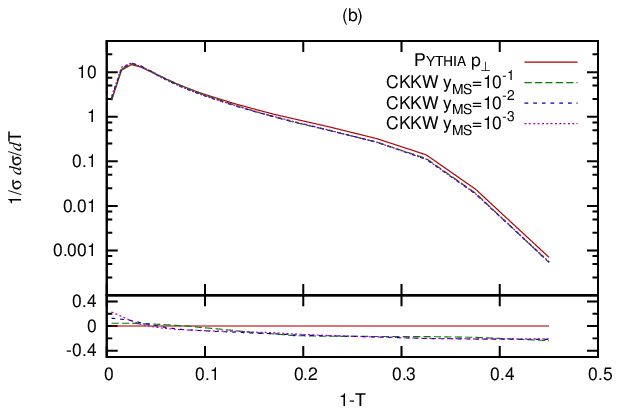,width=0.5\textwidth}
    \epsfig{file=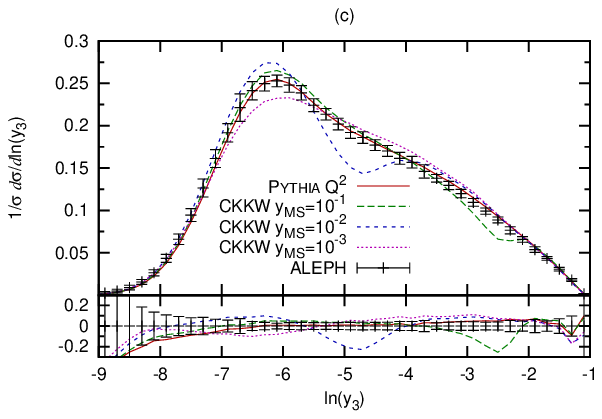,width=0.5\textwidth}%
    \epsfig{file=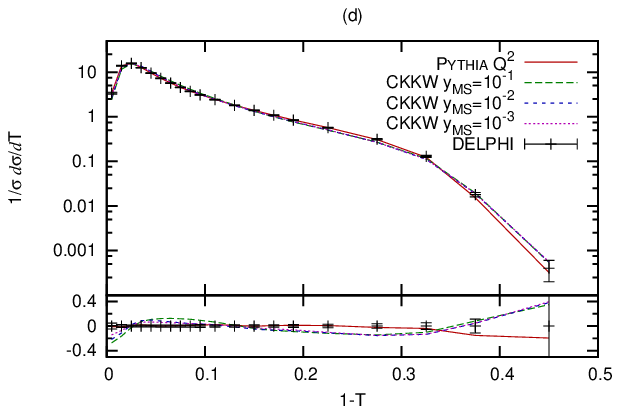,width=0.5\textwidth}
    \caption{\label{fig:ckkw-had} Charged plus neutral particle $y_3$ and
      charged particles thrust
      for our \protect\pythia implementation of CKKW corrections with
      transverse momentum and virtuality ordered showers and using
      different merging scales. Figure (c) includes a comparison to ALEPH
      data and (d) a comparison to DELPHI data, where figure (a) and (b)
      only compare to the default \protect\pythia, since the transverse
      momentum ordered shower has not yet been tuned to data.  }
}

Finally we show the consequences for two experimental observables in
figure \ref{fig:ckkw-had}. Again, the results of the \pT-ordered
shower has only been compared to the default \pythia, since it is not
yet tuned to data. The \pT-ordered plots show significant lower values
at high $y_3$ and low thrust, which is the result of the excess in
cross section for two-jet events. We also see some trace of the
discontinuities in figure \ref{fig:ckkw_py_max_y3}a, but they have
been smoothed out by the shower and hadronization. The results from
the virtuality ordred shower shows the same dips in the $y_3$
distribtion as in the parton-level plots. The dips are not visible in
the thrust plot, but there are significant deviations from default
\pythia.

\subsection{Pseudo-Shower}
\label{sec:res:pseudo}

We have implemented the Pseudo-Shower\cite{Mrenna:2003if} algorithm
using matrix element events generated with \madevent and using \pythia
with both virtuality ordered and $p_\perp$-ordered parton shower. The
implementation has been done with the following definition for the jet
clustering.
\begin{eqnarray}
\label{eq:luclus-kt}
d_{ij} \equiv m_{ij}^2 E_i E_j / (E_i + E_j)^2
\end{eqnarray}
This is equivalent to the measure in the \luclus
algorithm\footnote{Originally included in the \jetset program
  \cite{Sjostrand:1982am}, now a part of \pythia as the PYCLUS
  routine.} (and to the \kT-definition in \eqref{eq:pytpt}) in the
limit of massless particles. The jet algorithm was used both to define
the merging scale and the clustering of jet observables as specified
by the merging scheme. \as-reweighting has been introduced using the
scales from the jet clustering, which is happens to be the same scale
definition that is used in the \as\ evaluation in the \pythia shower.

\FIGURE[t]{
  \epsfig{file=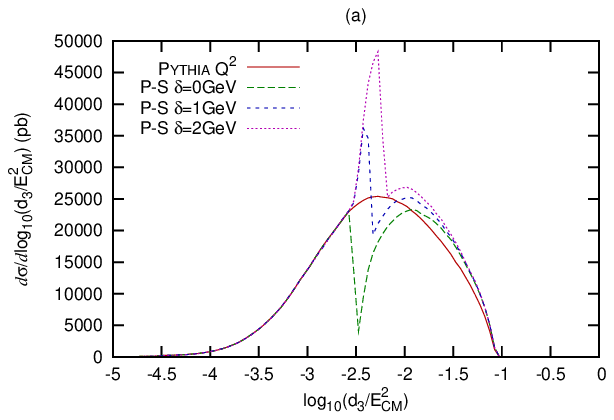,width=0.5\textwidth}%
  \epsfig{file=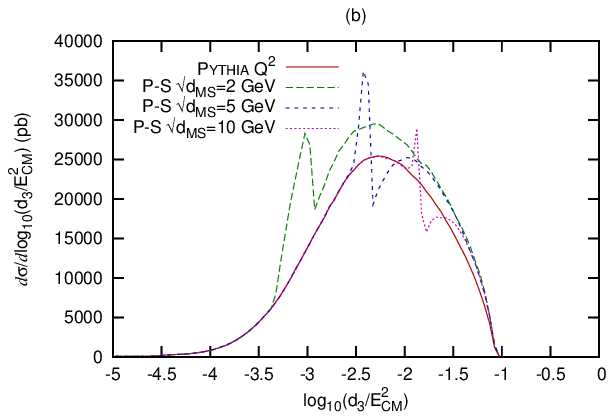,width=0.5\textwidth}
  \epsfig{file=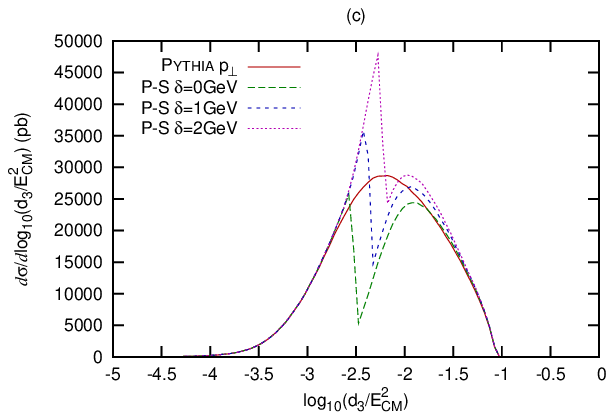,width=0.5\textwidth}%
  \epsfig{file=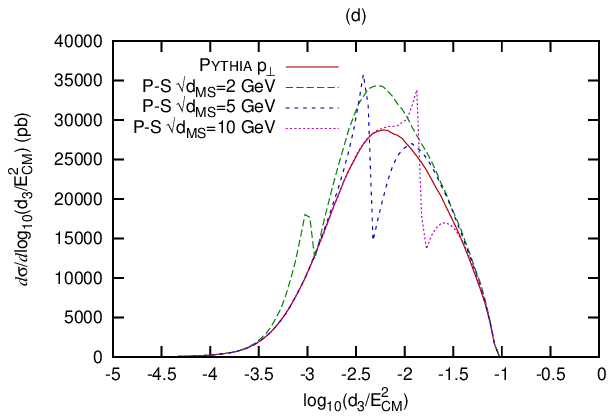,width=0.5\textwidth}
  \caption{\label{fig:ps_par_d3_dd} The $d_3$ spectra at parton level
    for \protect\pythia (with first order matrix element reweighting
    included) and for our implementation of the Pseudo-Shower
    algorithm using both the vituality ordered and the
    $p_\perp$-ordered shower. (a) and (c) show the effects of different
valuses of the fudge factor $\delta$ with a merging scale
of $\sqrt{\dms} = 5$ GeV. (b) and (d) show the results from different
values of the merging scale $\dms$ using a fudge factor $\delta=1$~GeV.}
}

The first thing that is investigated is the effects of different values
of the fudge factor $\delta$ introduced in
section~\ref{sec:pseudo-shower}. The most sensitive distribution for
checking the effects is the same variable as the merging scale. Figure
\ref{fig:ps_par_d3_dd}a and \ref{fig:ps_par_d3_dd}c show the $d_3$
distribution at parton level for three different values of $\delta$. In
both figures the curve that shows the smallest deviations is $\delta =
1$ GeV, and this value is used in the rest of this section. It is clear
that there is no smooth transition between the matrix element and parton
shower phase space.

Figure \ref{fig:ps_par_d3_dd} also showes what happens if the merging
scale is varied. The discontinuities persist for all three values of the
merging scale and for the two different showers and the results are heavily
dependent on the merging scale. The problem is a conseqence of using
different ways of defining the scales in the algorithm, which we
discussed in section \ref{sec:pseudo-shower}. The results in the rest of
this section focuses on the Pseudo-Shower implemenation with a
vituality ordered shower, since this was used in the original
publication \cite{Mrenna:2003if}.

\FIGURE[t]{
  \epsfig{file=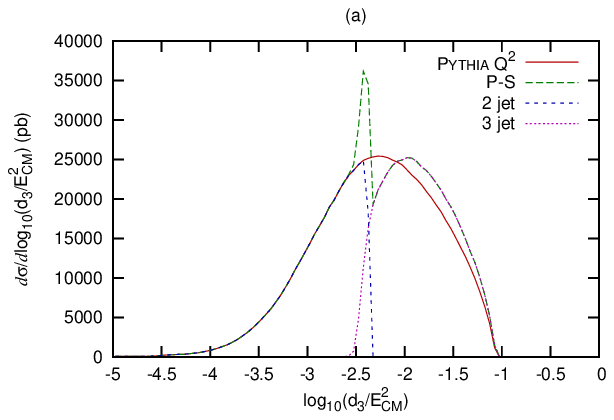,width=0.5\textwidth}%
  \epsfig{file=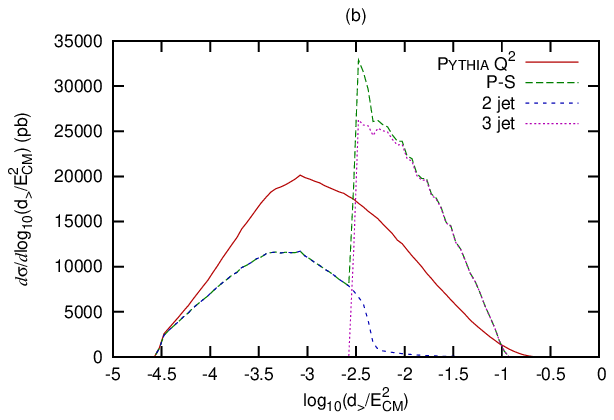,width=0.5\textwidth}
  \caption{\label{fig:ps_par_d3_j} The Pseudo-Shower results, with
    $\sqrt{\dms} = 5$ GeV and $\delta = 1$ GeV, split into two- and
    three-jet components compared to the default shower in \protect\pythia
    (with first order matrix element reweighting included). (a) shows the
    $d_3$ distribution at parton level and (b) is the largest $d$-value
    of the emissions in the shower.}
}

To demonstrate the different scales used in the algorihtm, the
Pseudo-Shower results have been split up into the contributions from the
two- and three-jet matrix elements. Figure \ref{fig:ps_par_d3_j}a shows
the $d_3$ distribution, where it is clear that the two-jet contribution
displays a sharp cut, but not the three-jet component. On the other
hand, when the largest $d$-value of the emissions in the shower is
plotted in figure \ref{fig:ps_par_d3_j}b, the three-jet component has a
sharp cut but not the two-jet curve. This illustrates that different
scales are used for the different jet multiplicities and this is the
reason for the problems that appear close to the merging scale.

\FIGURE[t]{
  \epsfig{file=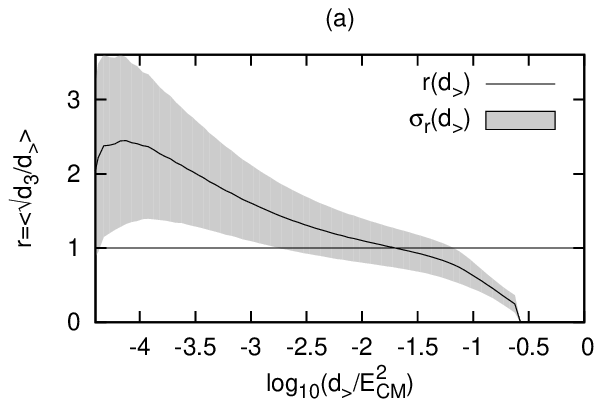,width=0.5\textwidth}%
  \epsfig{file=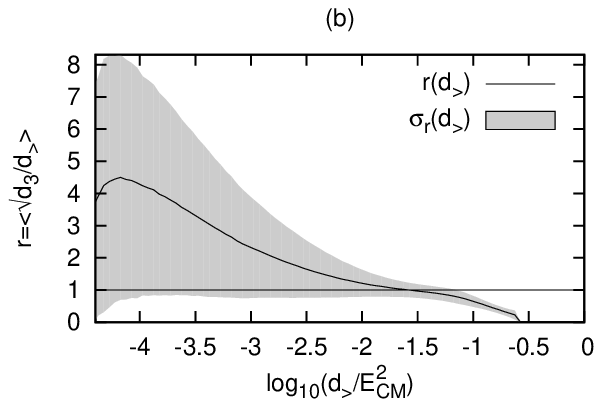,width=0.5\textwidth}
  \caption{\label{fig:d-ratio} The average ratio between the
    clustering scale $d_3$ for partonic jets and the generated maximum
    splitting scale $\rho_3$ in the cascade as a function of the
    latter. The shaded area indicates the standard deviation of the
    ratio. Both scales are defined as in \eqref{eq:luclus-kt}. (a) is
    for the $p_\perp$-ordered shower in \protect\pythia,
    while (b) is for the virtuality ordered one.}
}

To further illustrate the complication with mixing scales for parton
splittings and partonic jets, we show in figure \ref{fig:d-ratio} the
variation of the ratio between the $d_3$ scale on partonic jet level and
the largest generated scale in the parton shower $d_>$. In a strongly
ordered shower this ratio should ideally be unity, especially for the
\pT-ordered \pythia shower in figure \ref{fig:d-ratio}a. However, we
find that this is far from the case. Especially for the virtuality
ordered shower in figure \ref{fig:d-ratio}b, the correlation between the
different scales is very weak.

\FIGURE[t]{
  \epsfig{file=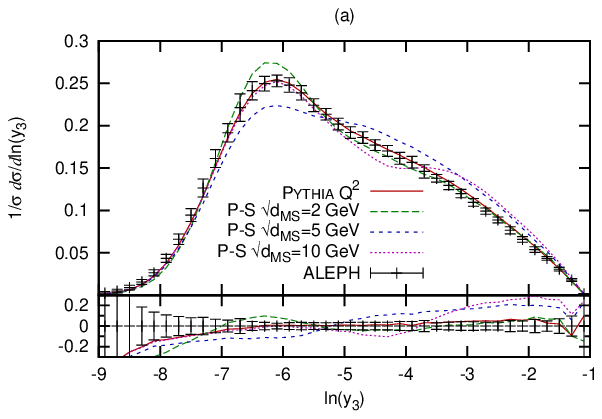,width=0.5\textwidth}%
  \epsfig{file=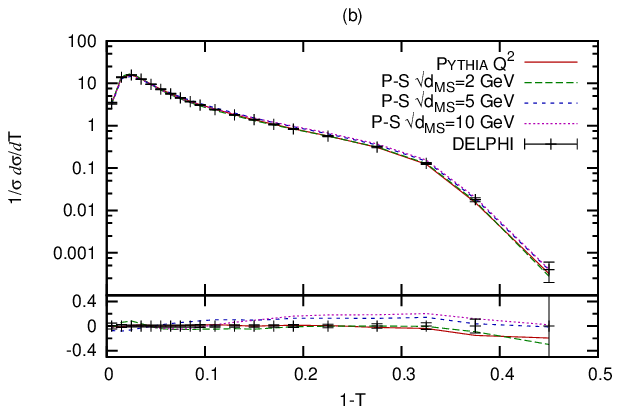,width=0.5\textwidth}
  \caption{\label{fig:ps_had_y3} The charged plus neutral particle $y_3$
    spectra (a) and charged particle thrust spectra (b) for \protect\pythia
    (with first order matrix element reweighting included) and for our
    implementation of the Pseudo-Shower algorithms with different
    merging scales and $\delta = 1$ GeV compared to ALEPH and DELPHI
    data.}
}

Finally two experimental observables are plotted, namely the normalized
$y_3$ distribution for charged and neutral particles and the charged
particles thrust, which are shown in figure \ref{fig:ps_had_y3}.
Hadronization smoothes out most of the discontinuities shown in earlier
figures, but it is clear that there are still problems. The plots show
deviations from data and the results are not be independent of the
merging scale.

\subsection{MLM}

To test MLM for $e^+e^-$ a merging scheme and a parton--jet distance
need to be chosen. Most of the MLM implementations use cone algorithms
to achieve this purpose. In the case of $e^+e^-$, we think that a
\kT-based clustering algorithm is a better choice. We have therefore
decided to use the Durham \kT\ algorithm to define the merging scale
and matrix element cutoff.

When it comes to selecting a measure for the parton--jet distance, the
natural choice would be the \kT-distance between the jet and the parton.
The problem with this approach is that since the \kT-measure includes the
minimum of the two energies it means that soft partons can match jets at
very wide angles, which leads to problems with convergence. Instead we
modified the distance to only use the jet energy.
\begin{eqnarray}
y_{\mrm{jet,parton}} = 2E_{jet}^2 (1 -
\cos(\theta_{\mrm{jet,parton}})) / E_{CM}^2
\end{eqnarray}

For the highest multiplicity treatment we follow the \madevent
implementation in \cite{Alwall:2007fs}. For each highest multiplicity
event, jets are reconstructed at a scale which is the maximum of the
merging scale and the smallest distance between the partons in the
matrix element level event, $\max(\yms, y_N)$. This scale is also used
when matching the jets to partons. This allows extra jets to be
produced if they are softer than those from the matrix element.

The algorithm has been implemented using \madevent to generate
the matrix element event and both \herwig and \pythia has been
used to shower the events. Before we move on to the results of
the merging, some important aspects of the showers need to be
explained.

Step \ref{step:mlm_shower} in the algorithm (described in section
\ref{sec:theory_mlm}), where the shower is invoked, depends on how the
state received from the matrix element is treated in the parton shower
program. For the analysis of the merging it is important to understand
how the events are treated internally. In particular the states
containing soft and collinear partons require extra scrutiny since the
matrix element cross section is divergent in these regions.

\FIGURE[t]{
  \epsfig{file=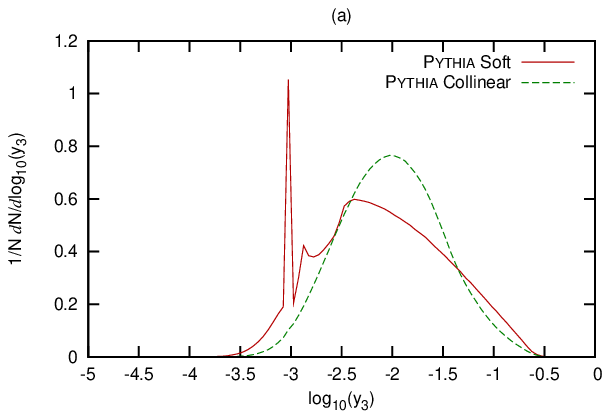,width=0.5\textwidth}%
  \epsfig{file=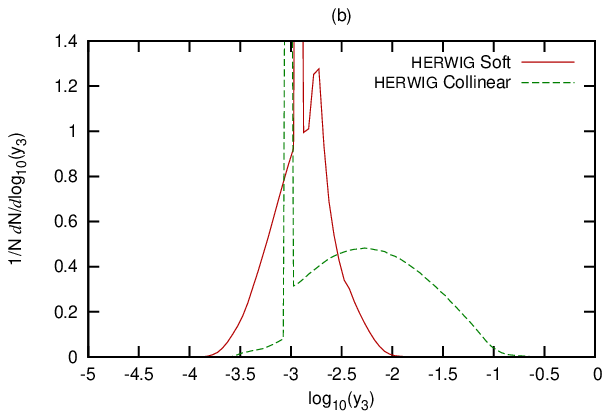,width=0.5\textwidth}
  \caption{\label{fig:fix_y3_pythia} The $y_3$ spectra at parton level
    for \protect\pythia (a) and \protect\herwig (b) when starting from
    a state with a soft or a collinear gluon with a merging scale of
    $y = 10^{-3}$ and the scale in the Les Houches interface set to
    $\ECM$.}
}

In the \pythia implementation of the Les Houches interface, the limiting
factors for the radiation are the scale from the Les Houches interface,
which is used as a veto, and the energy of the partons, which become the
maximum kinematically allowed value for the invariant mass. No extra
vetoes are applied for emissions at a wide angles. This means that even
when soft or collinear partons are present some events have a lot of
emissions, assuming that the scale in the Les Houches interface is set
to a high value. Figure \ref{fig:fix_y3_pythia}a shows the $y_3$ spectra
for a state with a soft gluon and for one with a collinear gluon with
$y_3 = 10^{-3}$ and the scale set to $\ECM$. It is obvious from the
figure that a soft or collinear parton does not limit the emissions
significantly.

\herwig uses a different strategy for the internal choices of scale.
Each parton is given a starting scale which is equal to the product of
the four-momentum of itself and its color neighbour (if a parton has two
color neighbours one is chosen at random). The parton is then boosted to
the frame where it is at a right angle compared to its color neighbour
and the shower invoked. Any scale set in the Les Houches interface is
included as a veto on the transverse momentum (approximately given by
\eqref{eq:luclus-kt}) of the emissions. The $y_3$ histogram from a
\herwig shower are shown in figure \ref{fig:fix_y3_pythia}b starting
from the same states and from the same scale as in figure
\ref{fig:fix_y3_pythia}a.  From the figure one can tell that a soft
parton limits the emission from the shower quite drastically, but for a
collinear parton about half of the events show significant jet activity.

Clearly, the way soft and collinear events from the matrix element
give rise to hard emissions influences the jet matching veto,
which is assumed to give the Sudakov suppression of these events, and
the different choices of scales in the shower can have a big impact on
these Sudakov form factors. This means that one has to be very careful
with generalizing the MLM algorithm to different showers.

\FIGURE[t]{
  \epsfig{file=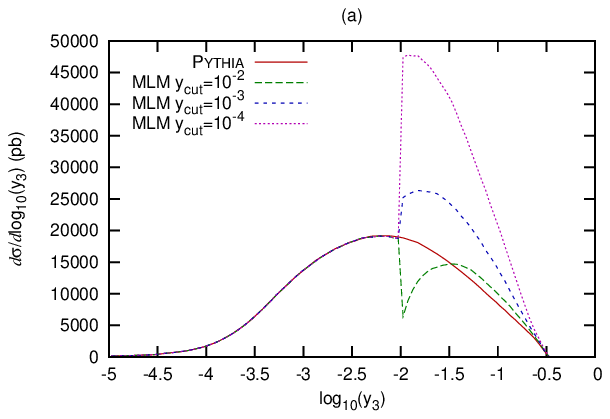,width=0.5\textwidth}%
  \epsfig{file=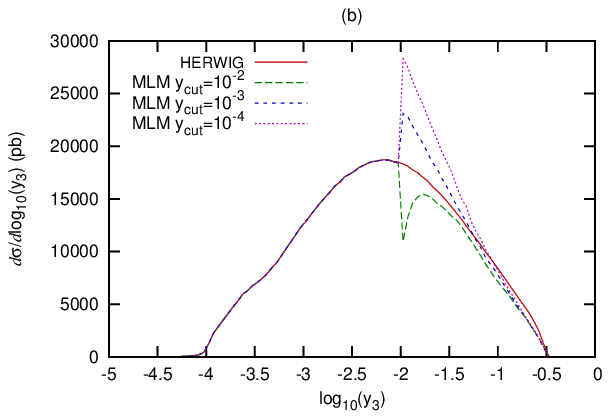,width=0.5\textwidth}
  \caption{\label{fig:mlm_par_y3} (a) The $y_3$ spectra at
    parton level for default \protect\pythia and our \protect\pythia
    implementation of the MLM algorithm for $y\sub{MS}=10^{-2}$ and for
    different values of the matrix element cutoff. (b) shows the same
    for \protect\herwig.}
}

The first thing that needs to be verified with our MLM implementation
is if the results converge when the matrix element cutoff is
lowered. The results of the merging are shown in figure
\ref{fig:mlm_par_y3} for \pythia and \herwig, with a merging
scale of $y_3 = 10^{-2}$ and three different values for the
cutoff. No sign of convergence is visible in the figure.

The problem is related to how \herwig and \pythia treats input with
soft or collinear partons. The MLM algorithm assumes that soft
and collinear partons cannot produce independent jets when fed
into the shower. From the earlier figures it is clear that \pythia
can produce extra jets both for soft and collinear partons,
whereas \herwig can only produce extra jets for collinear partons.
This is the reason why \pythia diverges faster.

\FIGURE[t]{
  \epsfig{file=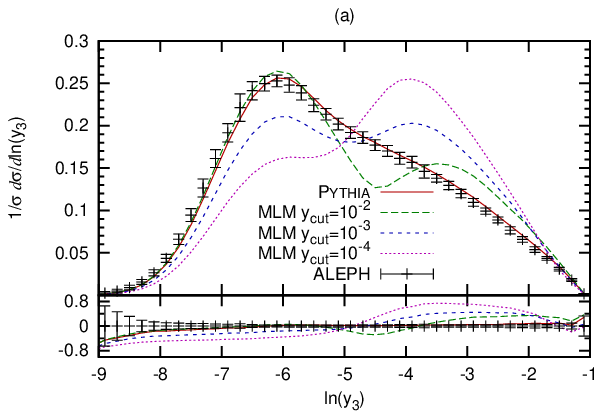,width=0.5\textwidth}%
  \epsfig{file=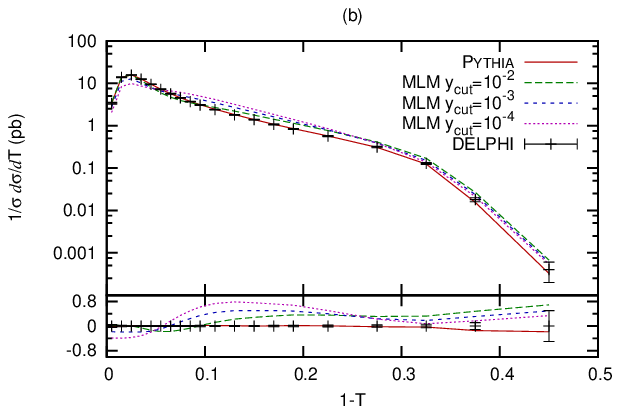,width=0.5\textwidth}
  \epsfig{file=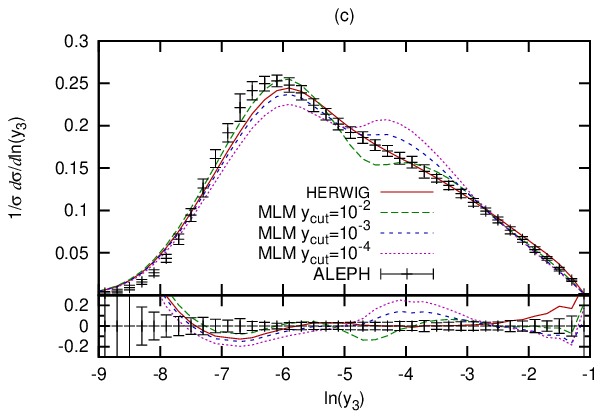,width=0.5\textwidth}%
  \epsfig{file=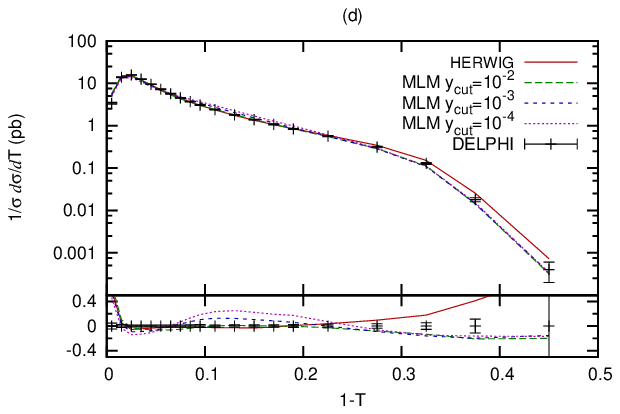,width=0.5\textwidth}
  \caption{\label{fig:mlm_had} The charged plus neutral particle $y_3$
    spectra (a) and the charged particle thrust spectra (b) for default
    \protect\pythia and our \protect\pythia implementation of the MLM
    algorithm for three different values of the matrix element
    cutoff. The \protect\herwig results are shown in (c) and (d)
    respectively.}
}

The changes to the parton-level spectra is also visible in the
hadron-level observables, shown in figure \ref{fig:mlm_had}.
It is clear that changing the matrix element cutoff also affects the
event shape, which diminishes the predictive power of the model.

\FIGURE[t]{
  \epsfig{file=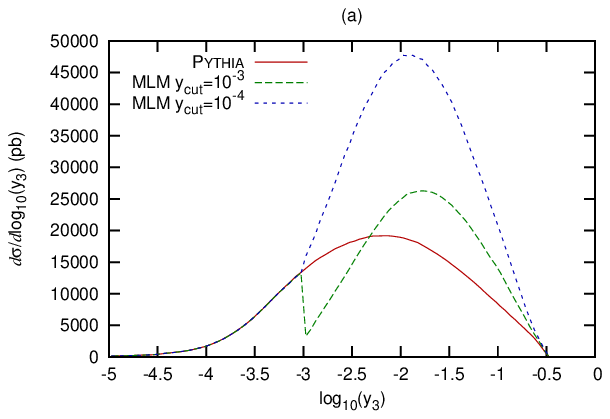,width=0.5\textwidth}%
  \epsfig{file=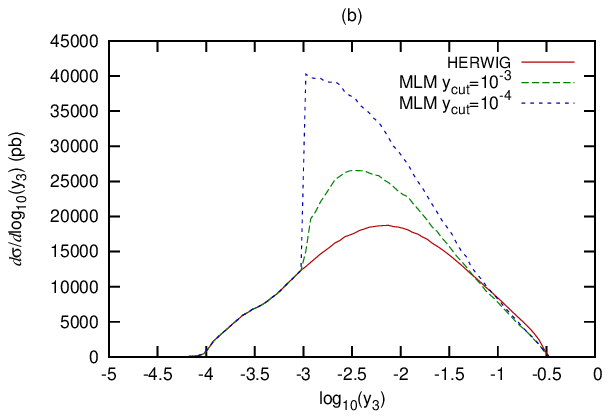,width=0.5\textwidth}
  \caption{\label{fig:mlm_par_y3_2} (a) The $y_3$ spectra at
    parton level for default \protect\pythia and our \protect\pythia
    implementation of the MLM algorithm for $y\sub{MS}=10^{-3}$ and for
    two different values of the matrix element cutoff. (b) shows the
    same for \protect\herwig.}
}

In figure \ref{fig:mlm_par_y3_2} the parton-level $y_3$-spectra is shown
with a lower merging scale. Generally one can say that the discrepancies
as compared to the standard \pythia and \herwig become even bigger. In
\herwig, configurations with soft gluons cannot generate hard emissions,
which means a large fraction of these events are kept. There is a pole
in the soft gluon limit and the cross sections for these events are
therefore rather large, which leads to a continued rise of the cross
section of the MLM-corrected curve for low values of $y_3$, clearly
visible in figure \ref{fig:mlm_par_y3_2}b.

The examples above show that the MLM algorithm is very sensitive to how
emission are generated and it appears to be quite tough to fix this
problem. If the shower generates an emission it risks influencing things
at higher jet scale and if emissions are suppressed it is not possible
to control the rise of the cross section near the soft and collinear
limits. In fact, even if the problems described in this section can
somehow be resolved, the more fundamental problem, described in section
\ref{sec:theory_mlm}, that MLM uses the wrong type of Sudakov form
factors remains unsolved.

\section{Conclusions}
\label{sec:conclusions}

In the previous sections, we have studied in some detail the behavior
of four suggested algorithms, or schemes, for merging fixed order,
tree-level matrix element generators with parton shower generators. We
have done so by considering the simplest possible case of the leading
order correction to the $\ee\to$ hadrons process. This may not be an
important use case for these merging schemes, but since it is such a
simple process it is fairly easy to check whether the schemes actually
accomplishes what they set out to do, namely to correctly populate the
phase space above the merging scale with partons described by the full
matrix element, and with emission from the parton shower below, and
that they do so while correctly resumming large logarithmic
contributions from soft and collinear divergencies. In addition, for
this process we also have the ``correct'' answer available, obtained
by a simple reweighting of the hardest emission in the parton shower.

Our main finding is that of the four schemes considered, only CKKW and
CKKW-L meets the requirements and that even CKKW has some problems
when using a parton shower with an ordering variable which is very
different from the clustering variable used for the merging scale.

For CKKW-L it was claimed in \cite{Lonnblad:2001iq} that the
cancellation of the dependence on the merging scale is complete when
using the \ariadne dipole shower. Here we show this explicitly and also
show that it is still true when using the transverse momentum ordered
shower in \pythia. This is not a completely trivial result, because it
involves an additional ambiguity in how to reconstruct the shower
history for \pythia which is not present in \ariadne.

In the CKKW case we also find a near complete cancellation of the
merging scale dependence when using the \pT-ordered \pythia shower.
However, when using the virtuality ordered shower in \pythia, we find a
clear mismatch near the merging scale, and we trace this mismatch to
different treatments of the no-emission probability from the gluon when
it is emitted above and below the merging scale, and from possible
problems in the parton shower when unordered emissions are allowed.
These problems occur when the ordering variable is very different from
the Durham \kT-measure used for the merging scale, which is true for the
virtuality ordering, but not for the \pT-ordered shower. We expect that
this is also the origin of the mismatches found in the CKKW
implementation in \herwig\cite{Mrenna:2003if}, where an elaborate tuning
of the scales used in the Sudakov and \as-reweighting was needed to
minimize the merging scale dependence. Most likely a similar retuning
can be done in the case of the virtuality ordered shower in \pythia.

For Mrennas Pseudo-Shower scheme, we find that the problems are much
more severe. The main problem is that cuts made on parton level for
the matrix element are made on the partonic jet level for the parton
shower, and we show that this can never give the clean separation of
phase spaces needed to have independence on the merging scale. This
leads to a severe underestimate of the three-jet rate just above
the merging scale. The fudge factor introduced in \cite{Mrenna:2003if}
can be tuned to hide the problem, but at the expense of overestimating
the three-jet rate below the merging scale due to double-counting.

The MLM scheme is the simplest one to implement, in that it allows the
use of any parton shower generator without modifying its internal
behavior. It also uses partonic jet level cuts, but contrary to the
Pseudo-Shower it tries to avoid mixing them with cuts on the few
parton level. Nevertheless, an extra parton-level generation cut is
needed for the matrix element. Supposedly the results should be
independent on this generation cut as long as it is sufficiently
smaller than the merging scale. However, we have found that this is
not the case, and that the result is very sensitive to how the
kinematics of the initial parton state limits emissions from the
chosen parton shower. If the emissions are not limited, events
generated close to the generation cut will always have a finite
possibility to end up above the merging scale, which makes the scheme
very sensitive to the generation cut. If, on the other hand, the
emissions are limited by the kinematics, there is no possibility to
obtain the necessary Sudakov suppression of soft and collinear
divergencies. It is not inconceivable that the generation cut can be
tuned, possibly together with the scale used in the \as-weighting, to
get reasonably smooth distributions, but we feel that this is just
hiding the fact that the MLM scheme has serious flaws.

Having done this investigation of the behavior of the algorithms for
the simplest possible process, the question is if we can make some
conclusions for more complicated processes. Or course, if a scheme
does not handle this simple process well, it is not likely that the
situation will improve for more complicated ones. But also in the case
of CKKW-L, complications may occur.

As we go to higher parton multiplicities in the matrix element, we
expect that the actual matrix element correction becomes larger,
necessarily giving rise to discontinuities near the merging
scale. These discontinuities should disappear as the merging scale is
lowered and the parton shower splittings become a good approximation
to the full matrix element. However, for this to work there must not
be any artificial dependencies on the merging scale in the merging
algorithm as such, which we have seen is not the case for all schemes.

The really interesting processes are in hadron collisions, where \eg\
the standard model production of a $W$ together with several hard jets
is an important background to almost any search for new
phenomena. Here the matching is complicated as the parton showers also
includes initial state evolution of the incoming partons. Also, care
must be taken to treat the parton densities in a consistent way. This
was first investigated for the CKKW scheme in \cite{Krauss:2002up},
and later for \ariadne implementation of CKKW-L in
\cite{Lavesson:2005xu}. For the latter it was shown that the merging
scale dependence does indeed cancel for the leading order correction
to $W$-production. Although it has not been checked explicitly for the
other schemes, there is no reason to expect that the problems we have
found in this paper will go away.

Having said all this, we must ask ourselves how severe the
deficiencies we have found here would be for practical applications at
\eg\ the LHC. We are, after all, dealing with tree-level matrix
elements and leading log parton showers which means we will in any
case expect large scale dependencies. And surely these deficiencies
results in much smaller uncertainties as compared to using parton
showers without matrix element corrections. Indeed, it was shown in
\cite{Alwall:2007fs} that CKKW, CKKW-L and different MLM
implementations all give consistent results for realistic experimental
observables within reasonable variations of the scales used in \as\
and parton densities. In fact the most severe disagreement found was
for the \ariadne implementation of CKKW-L. However, this had nothing
to do with the merging scheme, but is a consequence of the radically
different treatment of the phase space available for initial-state
radiation in \ariadne as such, which includes a resummation of some
logarithms of $x$, not present in conventional initial-state showers.

In \cite{Catani:2001cc} a large effort was put into showing that the
dependence of the merging scale vanishes at next-to-leading
logarithmic accuracy for CKKW, at least for a shower which is ordered
in the Durham \kT\ variable. Here we have shown that for the three-jet
observables considered here, the dependence vanishes completely for
CKKW-L. For the Pseudo-shower, the situation is more difficult to
analyze. In the strongly ordered limit we argue that it is equivalent
to CKKW-L, which indicates that the dependence on the merging scale
should cancel, at least to leading double-logarithmic accuracy,
nevertheless the dependence is quite large in absolute numbers. For
MLM, it is clear that for the observables considered here, it is
questionable if it can reproduce the correct Sudakov form factors,
indicating a dependence on the merging scale (or rather the cutoff in
the matrix element generation) already on the leading
double-logarithmic level. We plan to return in a future publication with
a more formal investigation of the logarithmic accuracy of the merging
scales.

In absolute numbers, we have seen that the Pseudo-Shower and the MLM
schemes, and even CKKW for some parton showers, have problems with
correctly populating different regions of phase space. And even if
this can be smoothed out by introducing additional cuts on the matrix
element generation (for MLM), a fudge factor (for Pseudo-Shower) or
separately varying the scale in the \as\ and Sudakov reweighting (for
CKKW with \herwig\cite{Mrenna:2003if}) it necessarily means
introducing extra parameters which need to be tuned. Indeed, it is not
inconceivable that these extra parameters need to be tuned differently
for different processes and merging scales, and maybe even for
different observables. As a consequence, the predictability of the
models will be reduced.

\section*{Acknowledgments}

We thank Frank Krauss, Peter Richardson, Torbjörn Sjöstrand and Stephen
Mrenna for useful discussions. Work supported in part by the Marie
Curie research training network ``MCnet'' (contract number
MRTN-CT-2006-035606).

\appendix

\section{Different scale definitions in CKKW}
\label{sec:appendix}

To illustrate the origin of the discontinuities found for CKKW when
used with the virtuality-ordered shower in \pythia, we here look at
the somewhat artificial example of calculating the differential
exclusive three-jet rate, where we use the notation introduced in
section \ref{sec:theory_ckkw-l}.

Assume the parton shower has ordering in $\rho$, and has auxiliary
splitting variables, $\vec{x}$, and that the Durham \kT-measure can be
written as a function of these variables, $y=y(\rho,\vec{x})$. Now we
can write the differential exclusive three-jet rate above some
$\rho_0$ for the plain parton shower case:
\begin{equation}
  \frac{d\sigma_3\sup{PS}}{\sigma_0}=\Delta_{S_2}\sup{PS}(\rho_{\max},\rho)
  \Gamma_{S_2}\sup{PS}(\rho,\vec{x})
  \Delta_{S_3}\sup{PS}(\rho,\rho_0)d\rho\,d^k\vec{x}
\end{equation}
where $\Delta_{S_2}(\rho_{\max},\rho)$ is the Sudakov corresponding to
the no-emission probability from the two-parton state above $\rho$,
and $\Delta_{S_3}(\rho,\rho_0)$ is the no-emission probability from
the three-parton state below $\rho$. We can also write the
corresponding rate for the CKKW scheme:
\begin{eqnarray}
   \frac{d\sigma_3\sup{CKKW}}{\sigma_0}&\approx&\Delta_{S_2}\sup{CKKW}(y_{\max},y)
   \Gamma_{S_2}\sup{ME}(\rho,\vec{x})\times\nonumber\\
   &&\Delta_{S_2}\sup{PS}(\rho_{\max},\rho;<y)
   \Delta_{S_3}\sup{PS}(\rho,\rho_0;<y)d\rho\,d^k\vec{x}
\end{eqnarray}
where $\Delta_{S_2}\sup{CKKW}(y_{\max},y)$ is the analytic Sudakov
form factor used for reweighting, and
$\Delta_{S_3}\sup{PS}(\rho,\rho_0;<y)$ is the parton shower
no-emission probability from the three-parton state between $\rho$ and
$\rho_0$ excluding phase space region corresponding to the vetoed
emissions with Durham \kT\ above $y(\rho,\vec{x})$.
$\Delta_{S_2}\sup{PS}(\rho_{\max},\rho;<y)$ is the corresponding
no-emission probability for the two-parton state, which approximates
the no-emission probability for the three-parton state above the scale
$\rho$ where the gluon is not allowed to radiate --- an approximation
which is valid as long as the gluon is not too hard. Now the CKKW
Sudakov can be approximately written as the no-emission probability
\begin{eqnarray}
  \Delta_{S_2}\sup{CKKW}(y_{\max},y)&\approx&
  \Delta_{S_2}\sup{PS}(\rho_{\max},\rho_0;>y)
  =\Delta_{S_2}\sup{PS}(\rho_{\max},\rho;>y)
  \Delta_{S_2}\sup{PS}(\rho,\rho_0;>y),
\end{eqnarray}
where the $>y$ notation indicates that only emissions with Durham \kT\
above $y$ are considered. Also, we can write
\begin{equation}
  \Delta_{S_2}\sup{PS}(\rho_{\max},\rho)=
  \Delta_{S_2}\sup{PS}(\rho_{\max},\rho;>y)
  \Delta_{S_2}\sup{PS}(\rho_{\max},\rho;<y)
\end{equation}
and
\begin{equation}
  \Delta_{S_3}\sup{PS}(\rho,\rho_0)=
  \Delta_{S_3}\sup{PS}(\rho,\rho_0;>y)
  \Delta_{S_3}\sup{PS}(\rho,\rho_0;<y)
\end{equation}
We can now rewrite the two three-jet rates as
\begin{eqnarray}
  \frac{d\sigma_3\sup{PS}}{\sigma_0}&=&
  \Gamma_{S_2}\sup{PS}(\rho,\vec{x})
  \Delta_{S_2}\sup{PS}(\rho_{\max},\rho;>y)
  \Delta_{S_2}\sup{PS}(\rho_{\max},\rho;<y)\times\nonumber\\
  &&\Delta_{S_3}\sup{PS}(\rho,\rho_0;>y)
  \Delta_{S_3}\sup{PS}(\rho,\rho_0;<y)d\rho\,d^k\vec{x}
  \nonumber\\
   \frac{d\sigma_3\sup{CKKW}}{\sigma_0}&=&\Gamma_{S_2}\sup{ME}(\rho,\vec{x})
   \Delta_{S_2}\sup{PS}(\rho_{\max},\rho;>y)
  \Delta_{S_2}\sup{PS}(\rho,\rho_0;>y)\times\nonumber\\
   &&\Delta_{S_2}\sup{PS}(\rho_{\max},\rho;<y)
   \Delta_{S_3}\sup{PS}(\rho,\rho_0;<y)d\rho\,d^k\vec{x}
\end{eqnarray}
And we find that the only difference, besides the desired
$\Gamma_{S_2}\sup{PS}\to\Gamma_{S_2}\sup{ME}$, is that in the plain
shower has a no-emission probability
$\Delta_{S_3}\sup{PS}(\rho,\rho_0;>y)$ where CKKW has
$\Delta_{S_2}\sup{PS}(\rho,\rho_0;>y)$. This can be seen as an extra
suppression in the plain parton shower due to the no-emission
probability from the gluon at a shower scale below $\rho$ but at a
Durham \kT-scale above $y$. Now if the ordering variable $\rho$ is
close to the Durham \kT, this region will be very small and the
mismatch will be small, as we saw in the case of the \pT-ordered
\pythia shower. But for the virtuality ordering in \pythia we have
\begin{equation}
  y=\min\left(\frac{z}{1-z},\frac{1-z}{z}\right)\frac{Q^2}{\ECM^2}
\end{equation}
and similarly for the angular ordering variable, $\xi$, in \herwig we
have
\begin{equation}
  y=\min(z^2,(1-z)^2)\frac{E^2\xi}{\ECM^2},
\end{equation}
(where $E$ is the energy of the parent parton in the splitting) and,
hence, the region can become quite large, especially in the regions
$z\to0$ or $1$, where the probability of an emission is large.  We
note, however, that in \pythia the effect is limited by the presence
of an additional angular ordering veto in addition to the virtuality
ordering.

\bibliographystyle{utcaps}
\bibliography{/home/shakespeare/people/leif/personal/lib/tex/bib/references,refs}

\end{document}